\documentclass[lettersize,journal]{IEEEtran}
\usepackage{amsmath,amsfonts}
\usepackage{algorithmic}
\usepackage{algorithm}
\usepackage{array}
\usepackage[caption=false,font=normalsize,labelfont=sf,textfont=sf]{subfig}
\usepackage{textcomp}
\usepackage{stfloats}
\usepackage{url}
\usepackage{verbatim}
\usepackage{graphicx}
\usepackage{cite}
\usepackage{hyperref}
\usepackage{cleveref}
\usepackage{float}
\usepackage{subfig}
\usepackage{amsmath}
\newtheorem{Def}{Definition}
\usepackage{listings}
\usepackage{xcolor}
\definecolor{USTCblue}{cmyk}{1,0.8,0,0}
\definecolor{Gaypurple}{RGB}{154,40,177} 
\definecolor{CommentGreen}{RGB}{0,139,0} 

\usepackage{multirow}
\usepackage{enumerate}
\usepackage{tcolorbox}
\tcbuselibrary{skins,breakable}
    {\endtcolorbox}

\usepackage{pifont}

\begin{document}
\bstctlcite{IEEEexample:BSTcontrol}

\title{NicePIM: Design Space Exploration for Processing-In-Memory DNN Accelerators with 3D-Stacked-DRAM}

\author{
        Junpeng Wang, % orcid 0000-0002-8810-3172 
        Mengke Ge, % orcid 0000-0001-7888-9370
        Bo Ding, % orcid 0000-0001-5939-7346
        Qi Xu, % orcid 0000-0002-0375-9800
        Song Chen, % orcid 0000-0003-0341-3428
        Yi Kang % orcid 0000-0002-5487-6855
        \thanks{\textbf{Preprint Notice: This work has been submitted to the IEEE for possible publication. Copyright may be transferred without notice, after which this version may no longer be accessible.}}
        \thanks{This work was supported in part by the National Key R\&D Program of China under grant No. 2019YFB2204800, in part by National Natural Science Foundation of China (NSFC) under grant Nos. 61931008, 62141415, U19A2074 and 61874102, in part by CAS Project for Young Scientists in Basic Research under grant No. YSBR-029k, in part by the Strategic Priority Research Program of Chinese Academy of Sciences, Grant No. XDB44000000. \textit{(Corresponding author: Song Chen.)}}
        \thanks{J.~Wang, B.~Ding and Q.~Xu are with School of Microelectronics, University of Science and Technology of China, Hefei, China. (e-mail: \href{mailto:wjp97@mail.ustc.edu.cn}{\texttt{wjp97@mail.ustc.edu.cn}}, \href{mailto:dingbo@mail.ustc.edu.cn}{\texttt{dingbo@mail.ustc.edu.cn}}, \href{mailto:xuqi@ustc.edu.cn}{\texttt{xuqi@ustc.edu.cn}})}
        \thanks{M.~Ge, S.~Chen and Y.~Kang are with School of Microelectronics, University of Science and Technology of China, Hefei, China and Institute of Artificial Intelligence, Hefei Comprehensive National Science Center. (e-mail: \href{mailto:mengke.ge@iai.ustc.edu.cn}{\texttt{mengke.ge@iai.ustc.edu.cn}}, \href{mailto:songch@ustc.edu.cn}{\texttt{songch@ustc.edu.cn}}, \href{mailto:ykang@ustc.edu.cn}{\texttt{ykang@ustc.edu.cn}})}
}

\markboth{}{Junpeng Wang et al.}


\maketitle

\begin{abstract}
With the widespread use of deep neural networks(DNNs) in intelligent systems, DNN accelerators with high performance and energy efficiency are greatly demanded. 
As one of the feasible processing-in-memory(PIM) architectures, 3D-stacked-DRAM-based PIM(DRAM-PIM) architecture enables large-capacity memory and low-cost memory access, which is a promising solution for DNN accelerators with better performance and energy efficiency. 
However, the low-cost characteristics of stacked DRAM and the distributed manner of memory access and data storing require us to rebalance the hardware design and DNN mapping. 
In this paper, we propose NicePIM to efficiently explore the design space of hardware architecture and DNN mapping of DRAM-PIM accelerators, which consists of three key components: PIM-Tuner, PIM-Mapper and Data-Scheduler. 
PIM-Tuner optimizes the hardware configurations leveraging a DNN model for classifying area-compliant architectures and a deep kernel learning model for identifying better hardware parameters. 
PIM-Mapper explores a variety of DNN mapping configurations, including parallelism between branches of DNN, DNN layer partitioning, DRAM capacity allocation and data layout pattern in DRAM to generate high-hardware-utilization DNN mapping schemes for various hardware configurations. 
The Data-Scheduler employs an integer-linear-programming-based data scheduling algorithm to alleviate the inter-PIM-node communication overhead of data-sharing brought by DNN layer partitioning. 
Experimental results demonstrate that NicePIM can optimize hardware configurations for DRAM-PIM systems effectively and can generate high-quality DNN mapping schemes with latency and energy cost reduced by 37\% and 28\% on average respectively compared to the baseline method. 
\end{abstract}

% Dear Editor in-chief, Associate Editor and Reviewers, 
% Thank you very much for reviewing our submission and providing your valuable comments. 

% We have revised this version by addressing reviewers, comments and incorporating their suggestions. All the recent changes in the current version are highlighted with blue color. Detailed responses to reviewers' comments are appended by the end of this manuscript. 
% Thank you very much. 
% Sincerely,  
% Junpeng Wang, Haitao Du, Bo Ding, Qi Xu, Song Chen, Yi Kang

\begin{IEEEkeywords}
Processing-in-memory, DNN accelerator, design space exploration.
\end{IEEEkeywords}

\section{Introduction}
\label{sec:intro}
% introduction to DNN. 
Deep neural networks(DNNs) have been used in many fields including image recognition, object detection and natural language processing, showing unprecedented accuracy. 
The majority of operations in DNNs are multiply-accumulate(MAC) operations with a large amount of data reuse, which makes DNNs compute-intensive and memory-intensive. 
% concerns of traditional DNN accelerators
With the scale of DNNs increasingly growing, the acceleration becomes a critical issue in the application of DNNs. 
Many domain-specific DNN accelerators have been proposed to get improved performance and energy efficiency\cite{chen_dadiannao_2014, du_shidiannao_2015,alwani_fused-layer_2016,chen_eyeriss_2016, jouppi_-datacenter_2017}.
Due to the large memory footprint of DNNs, one of the major concerns of these DNN accelerators is the costly off-chip DRAM access.
The memory hierarchy of DNN accelerators is elaborately designed to reduce off-chip DRAM access. A large part of the area of the chip is spent on buffers to make data more reused on chip. 
Elaborate scheduling strategies are often employed to make sufficient use of the capacity of the on-chip memory\cite{li_smartshuttle_2018,wei_overcoming_2019,chen_communication_2020}. 
% Off-chip memory access become more significant in the application of DNNs. 

\begin{figure}[htb]
    \centering
    \includegraphics[width=0.4\textwidth]{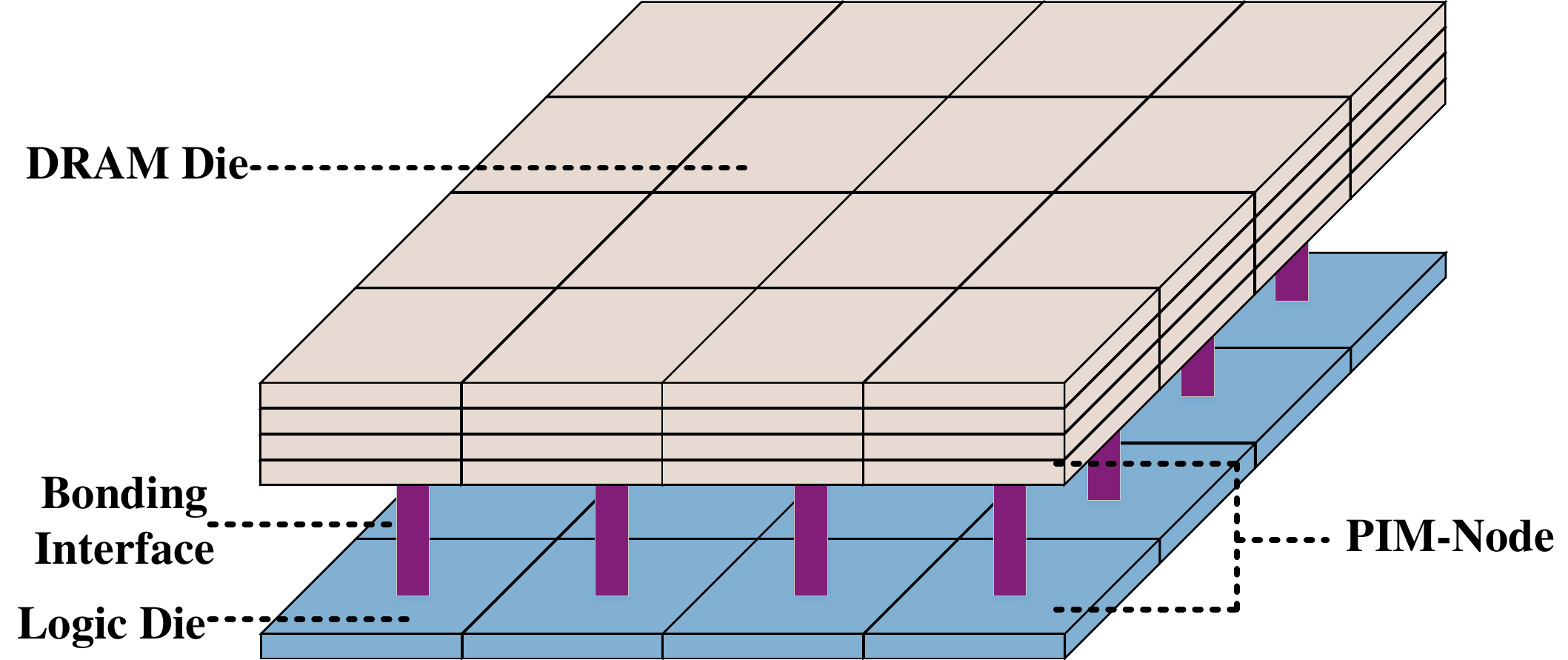}
    \caption{The PIM system with 3D-stacked DRAM. }
    \label{fig:PIM system overview}
\end{figure}

% introduction to 3D-stacked PIM
The technology of 3D-stacked memories enables the integration of large-capacity memory with low access cost\cite{hmc_2018, fujun_stacked_2020, shiba_96-mb_2021}, which provides a promising solution to the memory wall problem\cite{horowitz_11_2014}. 
In systems with 3D-stacked memory, the stacked logic die has the same area as the memory die and they are integrated by 3D-stacking technologies such as through silicon via(TSV)\cite{hmc_2018}, hybrid bonding\cite{fujun_stacked_2020}, etc. 
Among the widely used memory technologies, DRAM has relatively high density, so 3D-stacked-DRAM-based processing-in-memory system(DRAM-PIM system) is one of the promising choices for systems with high memory bandwidth and energy efficiency. 
The DRAM die contains an array of DRAM banks\cite{fujun_stacked_2020}(or vaults\cite{hmc_2018}) and each DRAM bank can be accessed independently in parallel instead of through the standard DDR interface. 
The 3D integration technology enables the stacked DRAM to have an order of magnitude higher bandwidth compared to conventional off-chip DRAM, and the closer distance between the memory and the logic makes the energy efficiency more than 5x better than off-chip DRAM\cite{fujun_stacked_2020}. 
3D-stacked DRAM has been used in many systems for the acceleration of memory-intensive applications \cite{kim_neurocube_2016, jiang_1596gbs_2021, niu_184qpsw_2022}. 
Due to the array-architecture of the DRAM, typically, as shown in \Cref{fig:PIM system overview}, the logic die is divided into an array and each part is combined with the corresponding DRAM bank(s) to form a function unit, which is denoted as a \textbf{PIM-node}. 
In each PIM-node, the logic part has independent access to the counterpart DRAM but each PIM-node has no direct access to the DRAM of other PIM-nodes. PIM-nodes communicate with each other through the on-chip-interconnect in the logic die such as network-on-chip(NoC). 
This distributed manner of computation and data storing benefits the utilization of the DRAM bandwidth of the DRAM-PIM system, but requires rebalancing the architecture design and the mapping algorithm. 

% recently proposed PIM accelerators. 
Recently proposed DRAM-PIM-based DNN accelerators are typically organized into a homogeneous tiled architecture\cite{kim_neurocube_2016,gao_tetris_2017,wang_exploiting_2017,wang_towards_2018,min_neuralhmc_2019}, which is in coordination with the array architecture of 3D-stacked DRAM. 
Each PIM-node can do DNN computation independently and  the NN engine in each PIM-node contains SRAM buffers and a processing-element(PE) array which can do MAC operations. 
The aforementioned works mainly focus on the architecture design and scheduling within one PIM-node, and they use simple DNN mapping strategies, paying little attention to the overhead brought by the distributed manner of computation and data storing. 
Besides, these works are customized designs for their workloads and targets, and may require manual tuning if the hardware constraints or the design target change. 
 
To meet the requirements of different targets under different hardware constraints, design space exploration(DSE) methods that generate high-quality hardware configurations and DNN mappings are necessary. 
The diverse choices of software mapping and hardware architecture of DRAM-PIM accelerators lead to a huge design space, making it impossible to find the optimal architecture by exhaustive search. In this work, given the configuration of the stacked DRAM with a certain number of DRAM banks and area, the following design space of DRAM-PIM accelerators is considered: \\
For \textbf{hardware configuration}, the granularity of PIM-nodes, PE array size and buffer sizes are taken into account: 
\textbf{(1)} For a DRAM-PIM system with a certain number of DRAM banks, larger but fewer PIM-nodes have fewer inter-PIM-node communication requirements while more but smaller PIM-nodes enable more mapping flexibility. The number of allocated DRAM banks for one PIM-node determines its DRAM bandwidth, DRAM capacity and area. 
\textbf{(2)}For one PIM-node with a certain number of allocated DRAM banks, the size of the PE array and sizes of SRAM buffers require a trade-off since a larger PE array increases the computing power and larger buffers allow more data reuse. \\
For \textbf{DNN mapping}, we consider the parallelism between branches of DNN, DNN layer partitioning, DRAM capacity allocation and data layout pattern in DRAM: 
\textbf{(1)} Many popular DNNs have multi-branch architecture such as multi-head-attention in Transformers\cite{devlin_bert_2019} and the inception block in GoogLeNet\cite{szegedy_going_2015}.
Making the branches processed in parallel rather than processing them serially on the PIM-node array can reduce the overhead brought by layer partitioning but may suffer from load imbalance between PIM-nodes. 
\textbf{(2)} A DNN layer needs to be partitioned so that it can be processed in parallel on multiple PIM-nodes. 
Different layer partition schemes correspond to different computation tasks of PIM-nodes and inter-PIM-node communication, which result in differences in performance. 
\textbf{(3)} The width of DRAM banks is larger than the data width of data of DNNs, especially when a PIM-node has many DRAM banks, thus proper data layout pattern in DRAM is required to achieve full use of dram bandwidth.  
\textbf{(4)} Due to the distributed manner of data storing, the DRAM of one PIM-node may not have enough capacity to store all weights of the whole DNN. If DRAM capacity is not sufficient for a PIM-node to store a whole replication of the weights of a layer, weights can be stored distributively and PIM-nodes share the weights when using. Sharing weights will require extra communication overhead while replicating the weights requires more DRAM capacity, so it is required to coordinate the weight replication values for all layers of the DNN. 

Existing design space exploration methods for DNN accelerators\cite{parashar_timeloop_2019,wu_accelergy_2019,venkatesan_magnet_2019, mei_zigzag_2021, zhang_full-stack_2022} have diverse prior definitions on the architecture for effectively searching for DNN mapping and efficiently selecting hardware configuration, and thus are not suitable for DRAM-PIM architectures with the aforementioned design space. 
% A design space exploration framework for DRAM-PIM systems is demanded to efficiently explore the design space and find high-quality legal architectures and corresponding high-hardware-utilization DNN mappings. 
In this paper, we propose a framework named NicePIM to optimize the hardware design and DNN mapping of DRAM-PIM-based DNN accelerators, and the main contributions of this paper are as follows:
\begin{enumerate}[(1)]
\item We propose NicePIM, a design space exploration framework for generating high-quality hardware design parameters and DNN mapping for DRAM-PIM-based DNN accelerators. NicePIM consists of a hardware design parameter optimizer(\textbf{PIM-Tuner}) that iteratively optimizes the hardware parameters and a DNN mapper(\textbf{PIM-Mapper}) with a \textbf{Data-Scheduler} to achieve high hardware utilization for various hardware configurations. 
% \item We propose a design space exploration framework consisting of a hardware design parameter optimizer(\textbf{PIM-Tuner}) that iteratively optimizes the hardware parameters and a DNN mapper(\textbf{PIM-Mapper}) with a \textbf{Data-Scheduler} to obtain high hardware utilization for various hardware configurations. 
\item The PIM-Tuner searches for better hardware configurations that make proper use of the limited area of the logic die of the DRAM-PIM accelerator, taking the granularity of PIM-nodes, size of PE array and sizes of buffers into account. PIM-Tuner consists of a DNN model for classifying area-compliant architectures and a deep kernel learning model\cite{wilson_deep_2015} for identifying hardware parameters with better quality. 
\item The PIM-Mapper explores a variety of DNN mapping configurations, including parallelism between branches of the DNN, DNN layer partitioning, DRAM capacity allocation  and data layout pattern in DRAM, for generating mapping schemes with high hardware utilization for various hardware configurations. 
\item To reduce the inter-PIM-node communication overhead of data-sharing due to DNN layer partitioning, the Data-Scheduler builds an integer linear programming(ILP) model to schedule the data transfer process, trying to balance the load of NoC links. 
\item Experimental results demonstrate that NicePIM can effectively optimize hardware configurations for DRAM-PIM systems and the proposed PIM-Mapper with the Data-Scheduler can reduce latency and energy cost by 37\% and 28\% on average respectively compared to the baseline method. 

\end{enumerate}

The remainder of this paper is organized as follows.
\Cref{sec:pre} presents preliminaries about DRAM-PIM systems and DNNs. 
\Cref{sec:design space} introduces the defined design space. 
\Cref{sec:overview} presents the overall flow of NicePIM with the following \Cref{sec:tuner}, \Cref{sec:mapper} and \Cref{sec:datascheduler} introducing the details of the PIM-Tuner, PIM-Mapper and Data-Scheduler respectively. 
The experimental results are shown in \Cref{sec:result}. 
\Cref{sec:related} lists some related works on 3D-stacked-memory-based PIM systems and design space exploration methods for DNN accelerators, followed by the conclusion in \Cref{sec:conclu}. 

\section{Preliminary}
\label{sec:pre}
\subsection{PIM systems based on 3D-stacked DRAM}
3D-stacked DRAM is a feasible solution for PIM systems with high performance and energy efficiency.
We use the 3D-stacked DRAM from UnilC\cite{fujun_stacked_2020} as the substrate of the PIM system in this work. 
In this architecture, the DRAM banks in the DRAM die are organized into an array architecture.
Each DRAM bank is connected to a controller in the corresponding part of the logic die and all controllers work independently. 
The function units of the DRAM-PIM system are placed in the remaining area of the logic die. 
Due to the array architecture of the DRAM banks, the function units in the logic die are divided into several parts and each part can directly access the DRAM bank(s) in the companion DRAM die. 
The function units and the corresponding DRAM bank(s) can be considered as an individual module denoted as a PIM-node. 
Each PIM-node accesses its own DRAM bank(s) with high speed and low cost but cannot directly access the DRAM banks of other PIM-nodes. 
We choose network-on-chip(NoC) as the on-chip-interconnect of the PIM-nodes for its feature of good extensibility and high-bandwidth. 

\subsection{DNN fundamentals}
\label{sec:fundamental}
A deep neural networks(DNN) consists of multiple layers to process data in a certain order.
The first layer receives the input data and the output of each layer is forwarded to the following layers according to the network topology.
Various kinds of layers are used in modern DNNs including convolution layer, matrix-multiplication layer(fully connected layer), pooling layer, normalization layer, etc.
A DNN may contain multiple kinds of layers but in most DNNs, convolution layers and matrix-multiplication layers account for the dominant part of the computation \cite{krizhevsky_imagenet_2017,he_deep_2016, simonyan_very_2015,szegedy_going_2015,redmon_yolov3_2018,bello_revisiting_2021}. 
\begin{figure}[htb]
    \centering
    \includegraphics[width=0.48\textwidth]{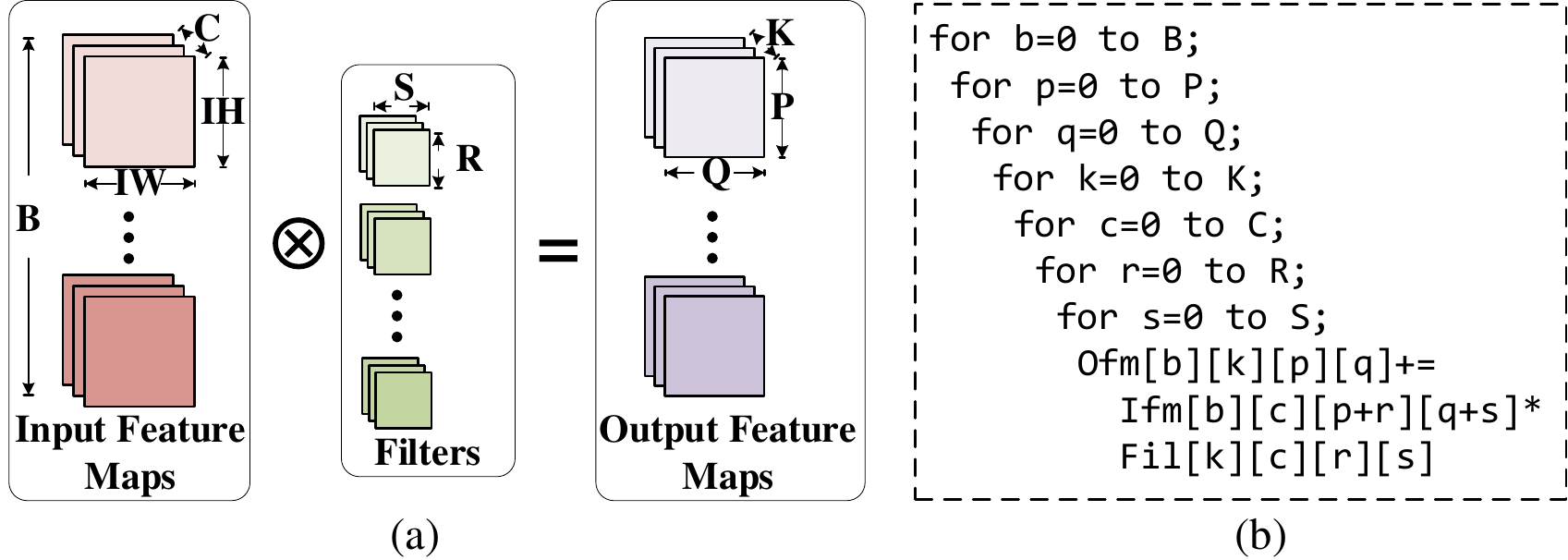}
    \caption{The computation process of a convolution layer. (a)filters slide on the input feature maps to generate output feature maps. (b)7 nested loops representing the computation process. }
    \label{fig:process of conv}
\end{figure}

A convolution layer has a set of filters that slide on the input feature maps(ifmaps) to generate the output feature maps(ofmaps), which is shown in \Cref{fig:process of conv}-(a). During the sliding process, a window of $C \times HK \times WK$, which is the same shape as the filter, is selected from the ifmaps and one point of the ofmap is generated after the dot-product of the selected window and the filter. 
For ifmaps of one sample in a batch, the sliding process of each filter is repeated for $P \times Q$ times. 
The computation process can be represented by the nested loops in \Cref{fig:process of conv}-(b). 
The feature maps generated by a convolution layer constitute a 4-D tensor, and we use $B$, $C$, $H$, $W$ to represent its batch size, number of channels, height and width, respectively. 
Most convolution layers are followed by activation functions to add non-linearity to the ofmaps. 

Matrix multiplication layers perform linear transformation for the inputs with the weight matrix.
This kind of layer multiplies the input matrix of dimension $B \times C$ with the weight matrix of dimension $C \times K$ to generate the output matrix with dimension $B \times K$.
Since the computation process can also be represented with the nested loops in \Cref{fig:process of conv}-(b) by setting the filter window size and ofmap size to $1\times 1$, in this paper, we use the representations of convolution layers to represent matrix multiplication layers for simplicity, including the loop dimensions and data dimensions. 

The topologies of DNNs are becoming more and more complicated. 
Most DNNs have linear structure and many popular DNNs have multi-branch architectures such as multi-head-attention in Transformers\cite{devlin_bert_2019} and the inception block in GoogLeNet\cite{szegedy_going_2015}. 
Many kinds of auxiliary layers are used to do down-sampling, concatenating, point-wise adding, point-wise multiplication, etc. These layers have simple computing processes and the number of operations is small so they are not major concerns in the design of DNN accelerators.

\section{Design Space Definition of the DRAM-PIM accelerator}
\label{sec:design space}
This section introduces the considered design factors for the DRAM-PIM accelerator. 
The hardware architecture and the hardware design parameters are introduced in \Cref{sec:hw design space}.
The following \Cref{sec:network level mapping}, \Cref{sec:layer level mapping}, \Cref{sec:DRAM capacity allocation} and \Cref{sec:data layout pattern} introduce the DNN mapping configurations that should be considered for getting high hardware utilization on DRAM-PIM accelerators with various hardware parameters. 
\subsection{Hardware configurations}
\label{sec:hw design space}
\begin{figure}[htb]
    \centering
    \includegraphics[width=0.48\textwidth]{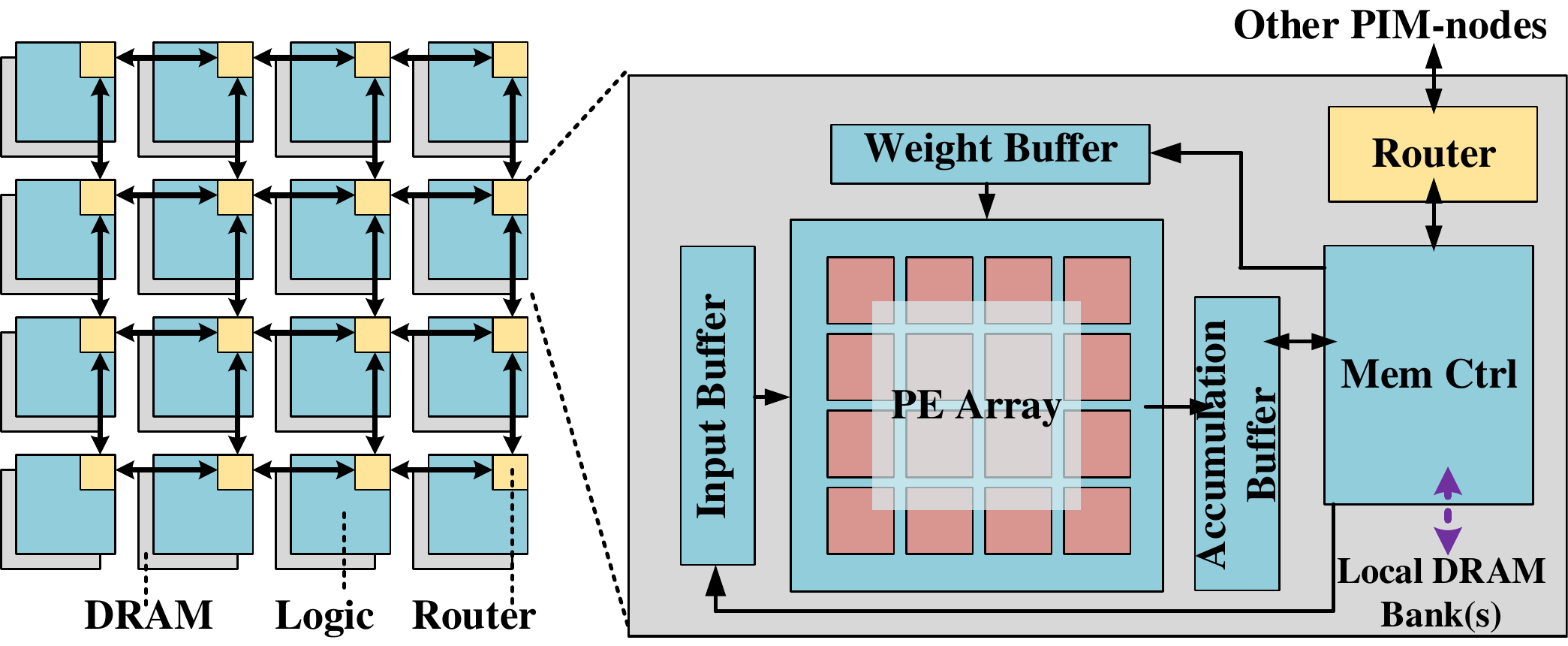}
    \caption{The architecture of a $4\times 4$ DRAM-PIM accelerator. }
    \label{fig:PIM accelrator}
\end{figure}
The hardware configuration mainly involves the granularity of PIM-nodes, PE array and buffers. 
As shown in \Cref{fig:PIM accelrator}, the DRAM-PIM system is a homogeneous 2-D PIM-node array, which is a widely used structure in many 3D-stacking-memory-based PIM systems\cite{kim_neurocube_2016,gao_tetris_2017,wang_towards_2018,ueyoshi_quest_2019}. 
A PIM-node consists of the stacked DRAM and the corresponding logic, and the logic part of the PIM-node has a NN engine, a DRAM bank controller and a router. 
PIM-nodes communicate with each other through the routers organized into mesh topology.  
A PIM-node can be allocated with one or several DRAM banks, and if a PIM-node has more than one DRAM bank, the ports of these DRAM banks are bound together so that the DRAM banks work in the same manner as one DRAM bank but with larger port-width. 
Besides, the total number of DRAM banks is constant so the number of allocated DRAM banks determines the number of PIM-nodes. 
The NN engine consists of a PE array and SRAM buffers for inputs, weights and outputs(and the partial-sums). 
The PE array is organized into parallel multiply-accumulation units, which is a widely used architecture \cite{nvdla,shao_simba_2019,venkatesan_magnet_2019}. 
% Loop $K$ and loop $C$ of convolution layers or matrix-multiplication layers are spatially mapped onto the PE array. 

The defined hardware design parameters are shown in \Cref{tab:HW design space}. 
We use the number of PIM-nodes to represent the granularity of PIM-nodes. 
The area, DRAM bandwidth and DRAM capacity of one PIM-node are proportional to the number of allocated DRAM banks. 
Larger but fewer PIM-nodes have fewer inter-PIM-node communication requirements since a DNN layer does not have to be partitioned into many parts to map on the PIM-node array. 
On the contrary, more but smaller PIM-nodes have more inter-PIM-node communication overhead, but enable more mapping flexibility. 
% More DRAM banks allocated to one PIM-node results in fewer PIM-nodes, which leads to less communication requirement between PIM-nodes but weaker mapping flexibility. 
For a PIM-node with a certain number of allocated DRAM banks, a larger PE array enables larger computing power, and increasing the sizes of buffers allows more data reuse. 
However, too large PE arrays and buffer sizes make the area dissatisfy the constraint. We have the following constraints for the hardware configurations: 
\begin{itemize}
\item[1)] The total area of the PIM-nodes should be no larger than the area of the DRAM die. 
\item[2)] $NA_{row}$ and $NA_{col}$ can exactly divide the rows and columns of the DRAM bank array to ensure homogeneous PIM-nodes. 
\end{itemize}
% https://www.tablesgenerator.com/
\begin{table}[htb]
    \caption{The hardware design parameters of the DRAM-PIM accelerator. }
    \label{tab:HW design space}
    \begin{tabular}{l|l|ll}
    \cline{1-3}
    Parameters & Type & Comment                                     &  \\ \cline{1-3}
    $NA_{row}$             & Int  & Number of rows of the PIM-node array            &  \\ \cline{1-3}
    $NA_{col}$             & Int  & Number of columns of the PIM-node array         &  \\ \cline{1-3}
    $PEA_{row}$            & Int  & Number of rows of the PE array of a PIM-node    &  \\ \cline{1-3}
    $PEA_{col}$            & Int  & Number of columns of the PE array of a PIM-node &  \\ \cline{1-3}
    $Size_{ibuf}$          & Int  & The input buffer size of a PIM-node             &  \\ \cline{1-3}
    $Size_{wbuf}$          & Int  & The weight buffer size of a PIM-node            &  \\ \cline{1-3}
    $Size_{obuf}$          & Int  & The accumulation buffer size of a PIM-node      &  \\ \cline{1-3}
\end{tabular}
\end{table}

\subsection{Parallelism between branches of DNN}
\label{sec:network level mapping}
\begin{figure}[htb]
    \centering
    \includegraphics[width=0.48\textwidth]{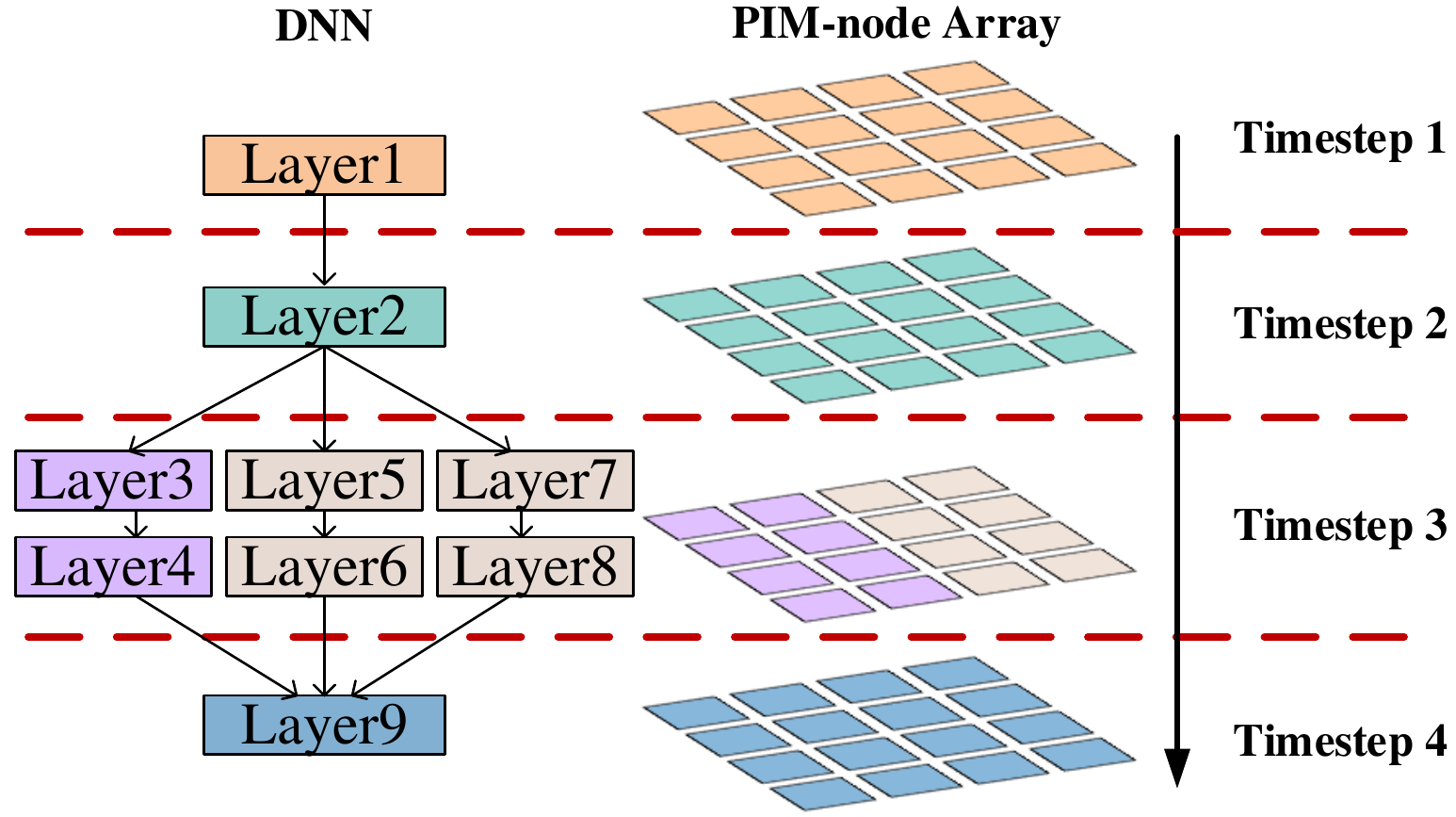}
    \caption{The example mapping of a DNN onto a $4\times 4$ PIM-node array. In each timestep, the layer is mapped onto the PIM-nodes with the same color. In timestep3, the PIM-node array is partitioned into 2 regions. One \textit{branch} of the 3-rd \textit{segment} is mapped onto the left part and the remaining two \textit{branches} are mapped to the right part. }
    \label{fig:DNN mapping}
\end{figure}

Since most DNNs have a linear structure and many popular DNNs have multi-branch architecture, we can make the branches processed on different regions of the PIM-node array to get inter-branch parallelism.  
We let the DRAM-PIM accelerator process the DNN in a timestep-by-timestep manner, which is shown in \Cref{fig:DNN mapping}. 
A DNN is partitioned into the smallest serial pieces possible and these parts of the DNN are denoted as \textit{segments}. 
The total number of \textit{segments} of a DNN is denoted as $N_{seg}$, which also means the DRAM-PIM accelerator requires $N_{seg}$ timesteps to process them.
If the $n_{seg}$-th \textit{segment} has a multi-branch structure, the different \textit{branches} can be processed in parallel and we denote the number of \textit{branches} as $N_{br}^{n_{seg}}$. 
Layers in one \textit{branch} are processed serially on the same region and for the $n_{br}^{n_{seg}}$-th \textit{branch}, we denote the number of layers as $L^{n_{br},n_{seg}}$. 
For example, in \Cref{fig:DNN mapping}, the $N_{br}$ of the 3-rd \textit{segment} is 3 and all \textit{branches} have 2 layers. 
% In each timestep, the PIM accelerator processes one \textit{segment}, which means if a DNN can be partitioned into $N_{seg}$ \textit{segments}, there are also $N_{seg}$ timesteps for the PIM accelerator. 
% For the $n_{seg}$-th \textit{segment}, we denote the number of \textit{branches} that can be processed in parallel as $N^{n_{seg}}_{br}$. Then in the $n_{seg}$-th timestep, we can partition the PIM-node array into at most $N^{n_{br}}_{br}$ regions and map the \textit{branches} onto them to gain layer-group-level parallelism. 

In the mapping process, we can make the \textit{branches} processed with different parallelism. 
Making more branches processed in parallel on the PIM-node array can reduce the overhead brought by layer partitioning but may suffer from load imbalance between PIM-nodes. 
For the $n_{seg}$-th \textit{segment}, the PIM-node array can be partitioned into at most $N_{br}^{n_{seg}}$ regions and we use $N_{reg}^{n_{seg}}$($1\leq N_{reg}^{n_{seg}}\leq N_{br}^{n_{seg}}$) to denote the number of regions. For the $n_{br}^{n_{seg}}$-th \textit{branch}, it can be mapped onto a region of  $n_{seg}$-th timestep and we denote the index of that region as $IR^{n_{br},n_{seg}}$ ($1\leq IR^{n_{br},n_{seg}}\leq N_{reg}^{n_{seg}}$). For example, in Timestep3 in \Cref{fig:DNN mapping},there are two regions and the $IR$ of the three \textit{branches} are 1, 2 and 2. 
In this work, we only consider mapping layers onto rectangular-shaped regions of the PIM-node array, so we use a position-shape pair, $((h_{pos}, w_{pos}), (h_{shape}, w_{shape}))^{n_{reg},n_{seg}}$, to represent the $n_{reg}^{n_{seg}}$-th region, $Region^{n_{reg},n_{seg}}$. 
The $h_{pos}$ and $w_{pos}$ indicate the smallest index on height and width dimension of the PIM-nodes of the region respectively, and the $h_{shape}$ and $w_{shape}$ indicate the height and width of the region of PIM-nodes respectively. 

In summary, the parameters for inter-branch parallelism of the $n_{seg}$-th \textit{segment} are:
(1) $N_{reg}^{n_{seg}}$, 
(2) $Region^{n_{reg},n_{seg}}$ and
(3) $IR^{n_{br},n_{seg}}$. 
Since for each \textit{segment}, the $N_{reg}$, $Region$ and $IR$ together determine the mapping, we use \textbf{SM}(\underline{S}egment \underline{M}apping Scheme) to represent them for simplicity. 

% The constraints of network level mapping is as follows: 

% 1) Hardware resource constraint: For each timestep, the regions of the PIM-node array should have no overlap. 

\subsection{Layer partitioning}
\label{sec:layer level mapping}
\begin{figure}[htb]
    \centering
    \includegraphics[width=0.48\textwidth]{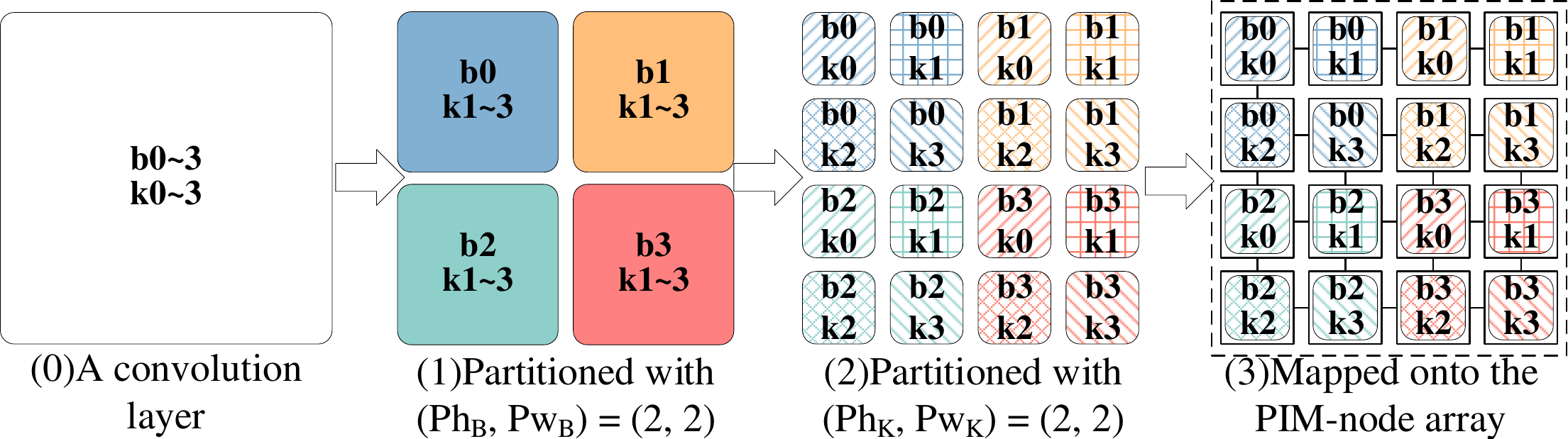}
    \caption{An example of the partitioning and mapping process of a convolution layer onto a $4\times 4$ PIM-node array. The non-1 partition numbers are $(Ph_{B}, Pw_{B})$ = (2, 2) and $(Ph_{K}, Pw_{K})$ = (2, 2) and the $P_{order}$ is $BPQKC$. The layer is firstly partitioned on the loop $B$ into $2\times 2$ part-layers (0)$\rightarrow$(1). Then each part-layer is partitioned on loop $K$ so that there are $4\times 4$ part-layers (1)$\rightarrow$(2). Finally, the part-layers are mapped onto the $4\times 4$ PIM-nodes correspondingly(2)$\rightarrow$(3). } 
    \label{fig:partition order example}
\end{figure}

A DNN layer needs to be partitioned so that it can be processed in parallel on the allocated PIM-node array\cite{kim_neurocube_2016, gao_tetris_2017}. Mapping a layer onto a $h_{shape}\times w_{shape}$ PIM-node array means the layer should be partitioned into $h_{shape}\times w_{shape}$ part-layers.
Different partition schemes result in different-shaped part-layers and different data requirements of the PIM-nodes, which influence the computing efficiency in each PIM-node and the amount of inter-PIM-node communication. 

To represent how the loops of a DNN layer are partitioned, we use five bi-tuples ($(Ph_{B}, Pw_{B})$, $(Ph_{P}, Pw_{P})$, $(Ph_{Q}, Pw_{Q})$, $(Ph_{K}, Pw_{K})$, $(Ph_{C}, Pw_{C})$) to denote the number of partitions for loop $B$, $P$, $Q$, $K$ and $C$, respectively. An example of layer partitioning is shown in \Cref{fig:partition order example}. For each loop of the layer, the loop length is divided by $Ph\times Pw$ to get the corresponding loop length of the part-layer. Loop $KH$ and $KW$ are not partitioned since they are relatively small. 

The spatial mapping of the part-layers determines the communication distance for transferring data of each PIM-node and thus influences the inter-PIM-node communication overhead. The order of spatial mapping of the part-layers, $P_{order}$, can be represented by a sequence of the loops $B$, $P$, $Q$, $K$ and $C$. An example of the spatial order is shown in \Cref{fig:partition order example}. 

For simplicity, we use \textbf{LM}(\underline{L}ayer \underline{M}apping Scheme) to denote the mapping scheme of a layer, which includes partition numbers ($(Ph_{B}, Pw_{B})$, $(Ph_{P}, Pw_{P})$, $(Ph_{Q}, Pw_{Q})$, $(Ph_{K}, Pw_{K})$, $(Ph_{C}, Pw_{C})$) and $P_{order}$. 

\subsection{DRAM capacity allocation}
\label{sec:DRAM capacity allocation}
The distributed data storing of DRAM-PIM systems makes the DRAM capacity of one PIM-node a constraint. 
Since NicePIM focuses on the inference process of DNNs, intermediate data can be discarded after being consumed while all the weights of the DNN should be stored in the DRAM-PIM system. 
The limited capacity of the DRAM may not be sufficient for one PIM-node to hold a whole replication of all weights of the DNN.

% If a layer is partitioned on the loop $B$, $H$ or $W$, the PIM-nodes that the layer is mapped require the same weights of the layer. 
% For $N_{node}$ PIM-nodes that require the same weights, if the DRAM capacity is not enough for each PIM-node to store a full replication of the weights, we make each PIM-node only store one part of the weight and PIM-nodes share the weights through NoC before processing the corresponding layer. 

If the DRAM capacity allocated for one layer is not enough for each PIM-node to store a full replication of the weights, we make each PIM-node only store one part of the weights and PIM-nodes share the weights through NoC before processing the layer. 
The weight-sharing process introduces extra inter-PIM-node communication cost. 
The less weight stored in PIM-nodes, the more required communication, which means we need to specify the number of weight replication for each layer. 
We use the number \textbf{WR}(\underline{W}eight \underline{R}eplication number) to represent the allowed number of replications of weights of one layer. 
%  if $WR\geq N_{node}$, each PIM-node can store a whole-replication of the required weights, which indicates no need to share weights. 
Denoting the number of PIM-nodes that require the same weights as $N_{node}$, if $WR$ is smaller than $N_{node}$, each PIM-node stores $\frac{1}{\lceil N_{node}/WR\rceil}$ part of the weights and the remainder parts are got from other PIM-nodes. 
Denoting the DRAM capacity of a PIM-node as $CAP$, for each PIM-node, the summation of the stored weights of all layers in that PIM-node should be smaller than $CAP$. 

\subsection{Data layout pattern in DRAM}
\label{sec:data layout pattern}
\begin{figure}[htb]
    \centering
    \includegraphics[width=0.48\textwidth]{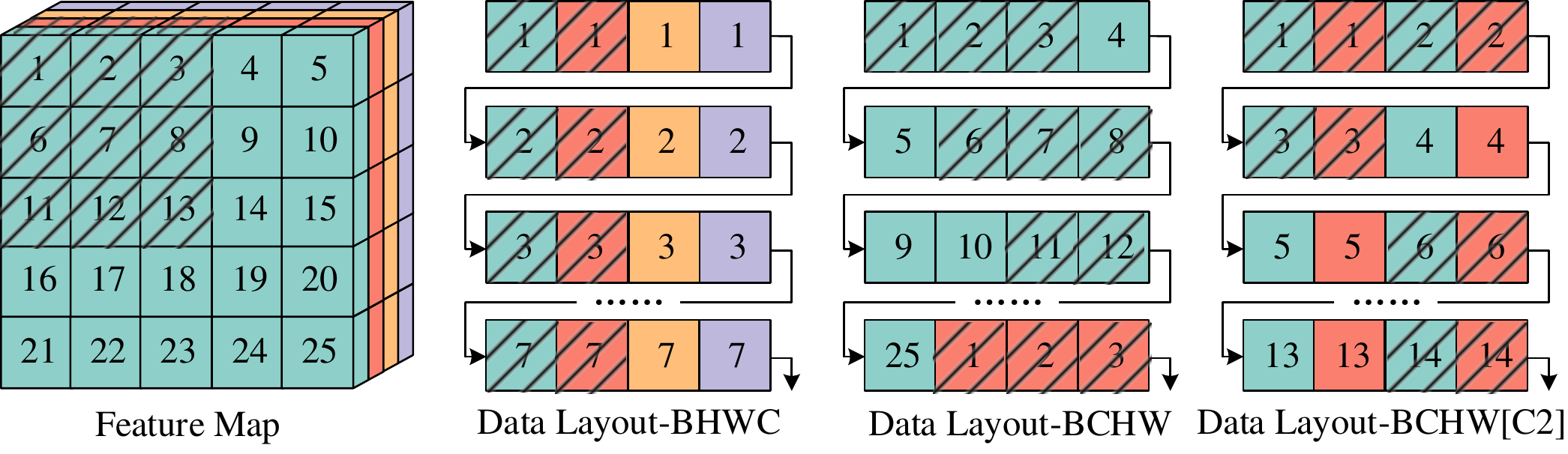}
    \caption{Four $5\times 5$ feature maps with different data layout patterns in DRAM, assuming four numbers per DRAM access. In the figure with \textit{BCHW}, the data is mapped in DRAM in the order of \textit{W-H-C-B}. In the figure with \textit{BCHW[C2]}, the data is firstly grouped every 2 channels and then mapped in DRAM. The two-channel $3\times 3$ window of the feature map covered with slashed lines illustrates the data access patterns with different data layout patterns. } 
    \label{fig:data layout pattern example}
\end{figure}
The pattern that the high-dimensional data of DNNs are flattened and stored in DRAM affects the efficiency of DRAM access. DRAM reaches its best performance and efficiency at sequential access since data access are performed via the row buffer. 
Row buffer miss or conflict will introduce extra energy and latency \cite{putra_romanet_2021} and the energy and latency cost of DRAM can be summarized as the summation of the cost of data access and row buffer updating. 
Besides, with DNN quantization widely used, the data width in DNNs is often much smaller than the width of DRAM banks(8/16-bit per data \textit{v.s.} 128bit per DRAM bank). 
If a PIM-node has more than one DRAM bank, the width difference between data and DRAM banks is even more critical. 
Since weights are read-only and can be re-arranged according to the access pattern in advance before being stored into the DRAM banks, we only focus on the data layout of input data and output data of DNN layers. 

To make DRAM access requirements more sequential, two data layout orders, \textit{BCHW} and \textit{BHWC}, are taken into account, which is illustrated in \Cref{fig:data layout pattern example}. 
To make full use of the width of DRAM banks, data grouping is employed before storing them into DRAM. 
An example of data grouping is shown in \Cref{fig:data layout pattern example} where the $[C2]$ indicates 2 channels of feature maps are grouped. 
If a $3\times 3$ window of the first two channels is selected to do convolution, the data layout with \textit{BCHW[C2]} requires 6 times DRAM access while the data layout with \textit{BCHW} and \textit{BHWC} requires 9 and 8 times respectively. 
We use $DL_{i}$ and $DL_{o}$(\underline{D}ata \underline{L}ayout Pattern) to represent the data grouping and layout order of the input and output data of a layer. 
The $DL_{i}$ and $DL_{o}$ of a layer can be different but for layers with data dependency, the $DL_{o}$ of the predecessor layer should be the same as the $DL_{i}$ of the successor layer since they stand for the same data. 
For simplicity, we use \textbf{DL}, which includes $DL_{i}$ and $DL_{o}$, to represent the data grouping and layout order of a DNN layer. 

\section{Overview of NicePIM}
\label{sec:overview}
\begin{figure}[htb]
    \centering
    \includegraphics[width=0.48\textwidth]{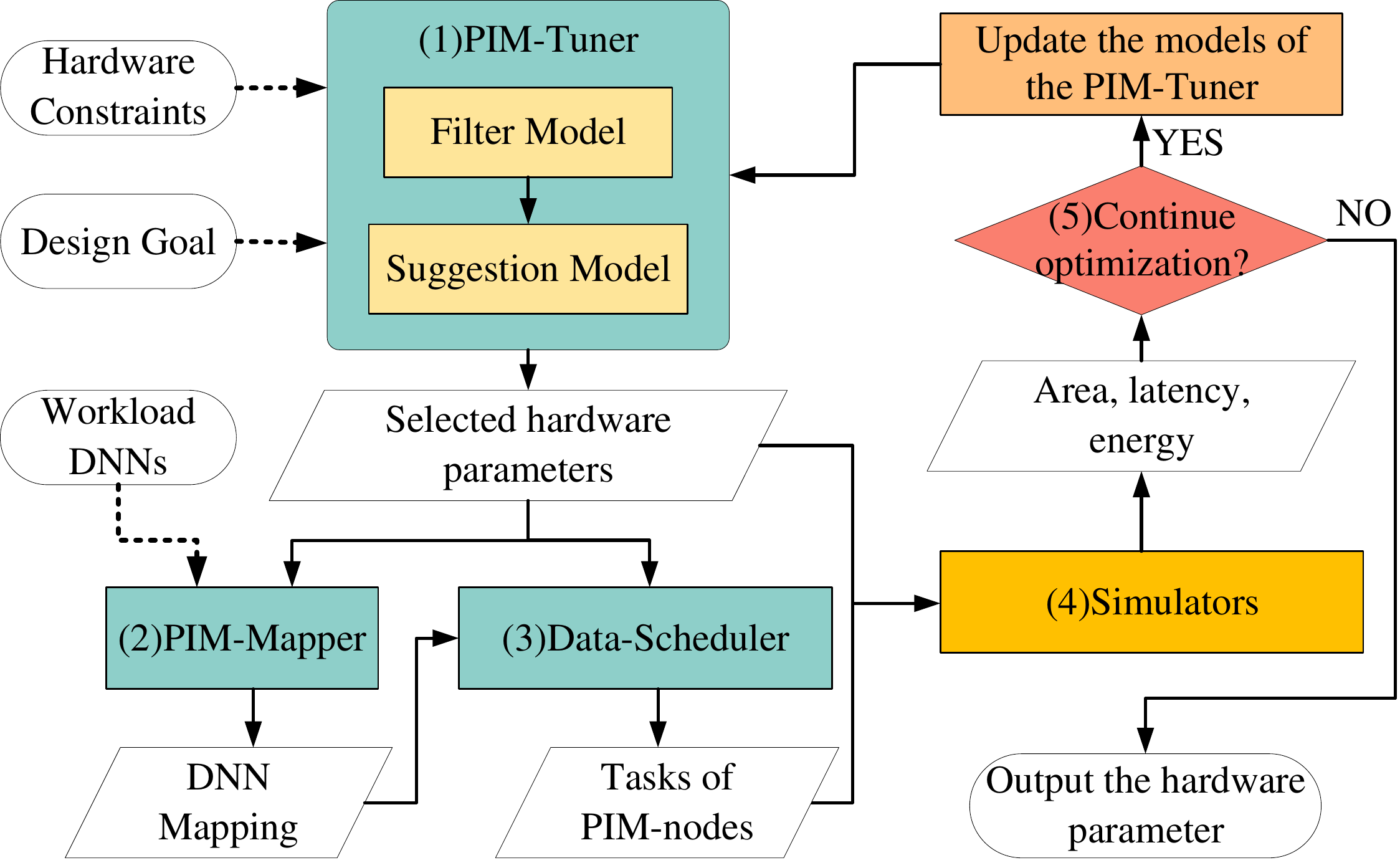}
    \caption{Overall flow of NicePIM. }
    \label{fig:overall flow}
\end{figure}
The overall flow of NicePIM is shown in \Cref{fig:overall flow}. 
The inputs of NicePIM include the hardware constraints, the design goal and the workload DNNs. 
The hardware constraints specify the attributes of the substrate of the hardware, such as the technology node, the total available area($Cstr_{area}$), the shape of the array of DRAM banks($BA_{row}\times BA_{col}$), the data width of each DRAM bank($Width_{bank}$), the capacity of one DRAM bank($CAP_{bank}$), etc.  
The design goal quantifies the hardware quality with given hardware design parameters, which can be expressed by a cost function related to energy and latency of each workload DNN: 
\begin{equation}
\label{eq:DesignGoal}
\begin{aligned}
    Cost = \sum_{DNN} Energy^\alpha \!\times\! Latency^\beta \times \gamma , \\ 
    \alpha \geq 0, \beta \geq 0, \gamma >0
\end{aligned}
\end{equation}
$\alpha$ and $\beta$ are to adjust the preference on latency and energy and $\gamma$ assigns different importance for each workload DNN. 

The design space exploration process of NicePIM is iterative, which is shown in \Cref{fig:overall flow}: 
\textbf{(1)} The PIM-Tuner samples a large batch of hardware parameters from the whole design space. 
Then hardware parameters that are predicted to exceed the area constraint according to the filter model are discarded. 
The remaining hardware parameters are sorted by the suggestion model so that the ones with better predicted performance are selected. 
\textbf{(2)}For each set of hardware parameters given by PIM-Tuner, the DNN mapper generates corresponding DNN mapping schemes for all workload DNNs. 
\textbf{(3)}Each mapping scheme produced by PIM-Mapper is translated into tasks of PIM-nodes, during which the data-sharing process is scheduled by the Data-Scheduler. 
\textbf{(4)}The selected architectures from PIM-Tuner are sent to the simulator to get the area one-by-one until one architecture with legal area is obtained. 
Then that architecture and the corresponding tasks of PIM-nodes are passed to the simulators to get the latency and energy of each workload DNN. 
\textbf{(5)}If the ending condition is not met, the simulation results of area, latency and energy are added to the datasets of the PIM-Tuner for updating its two models and then the iteration is repeated. 

\section{PIM-Tuner}
\label{sec:tuner}
\begin{figure}[htb]
    \centering
    \includegraphics[width=0.48\textwidth]{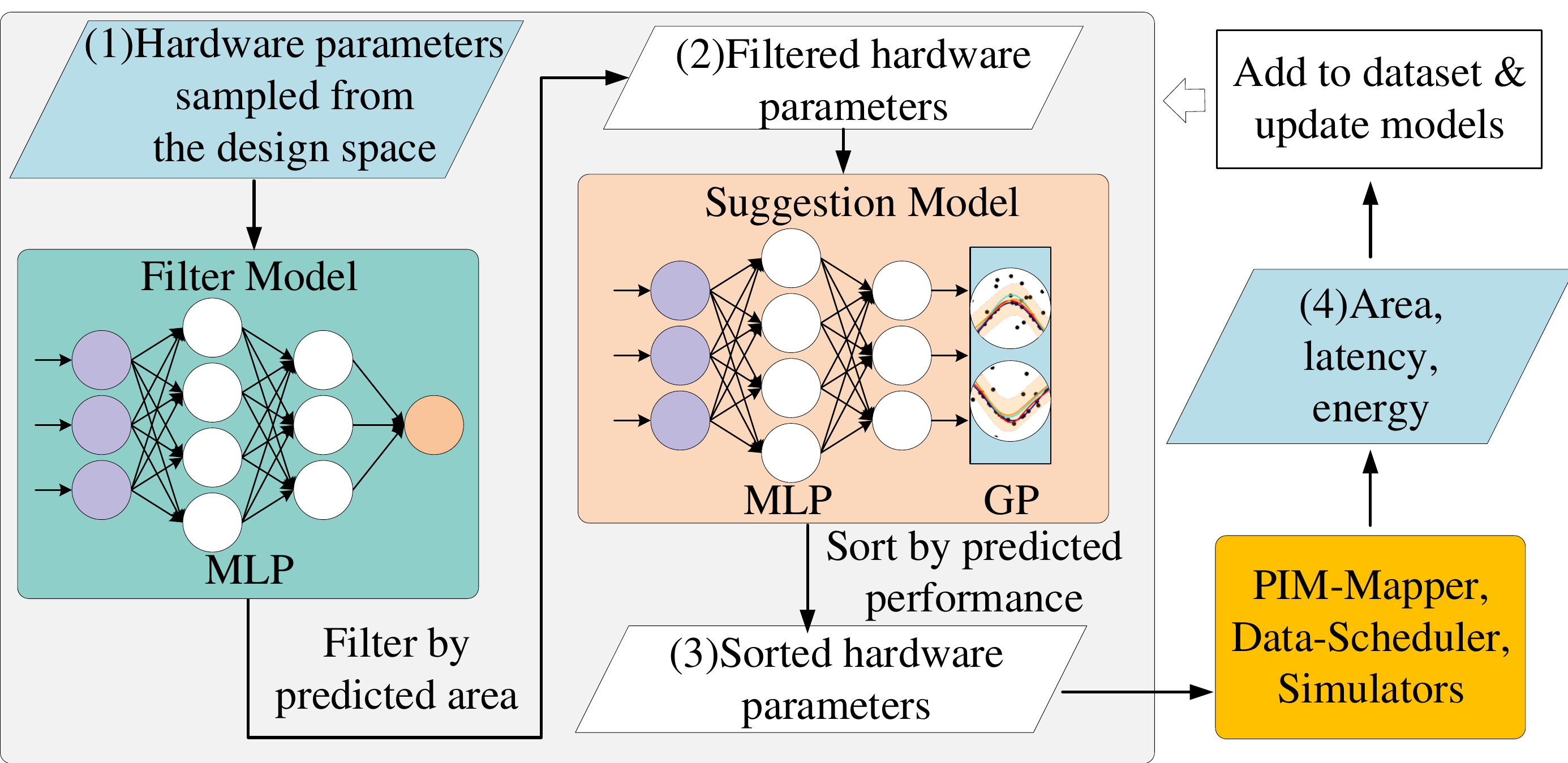}
    \caption{The flow of the PIM-Tuner in one iteration.}
    \label{fig:flow of tuner}
\end{figure}
The design space of hardware parameters illustrated in \Cref{sec:hw design space} is so large(for example, about $10^{10}$ in \Cref{tab:hardware constraints}) that it is infeasible to find the optimal one by enumerating all points in the whole design space. 
For effectively characterizing the complicated design space of hardware parameters, in PIM-Tuner, we build a filter model to predict the area and a suggestion model to identify better architectures. The flow of PIM-Tuner is shown in \Cref{fig:flow of tuner}. 
The simulation results from previous iterations of the design space exploration flow are collected to form datasets for updating the models. 
% We update the models once the dataset is updated at the end of each iteration of the design space exploration process. 
% Because of feature of Gaussian process models, we employ the Gaussian process model as the suggestion model.

The suggestion model is a deep kernel learning model\cite{wilson_deep_2015}, which combines the robustness and non-parametric flexibility of Gaussian process with the expressive power of deep learning models. 
The input of the suggestion model is the normalized vector of hardware parameters and the model is fitted with the costs of the architectures with the corresponding hardware parameters, which are calculated by the function in \Cref{eq:DesignGoal}. 
In the updating process of the suggestion model, we learn the Gaussian process model and the DNN model jointly by maximizing the log marginal likelihood of the Gaussian process. 

The DNN-based filter model is employed for identifying architectures that exceed the area constraint. 
Due to the 3D-stacking pattern of DRAM and logic, the area of the logic part of the DRAM-PIM accelerator is constrained by the DRAM part. 
Checking the area with simulators is time-consuming, so the filter model is necessary for reducing the times of invoking the simulators. 
The filter model gets hardware parameters as the input and outputs the corresponding area. 
We train the filter model using stochastic gradient descent algorithm with the mean squared error(MSE) between the output area and the simulated area as the loss function. 

% The flow of the PIM-Tuner in one iteration of the design space exploration process is shown in \Cref{fig:flow of tuner}. \textbf{(1)}Firstly, a batch of points in the hardware design space is randomly sampled. 
% \textbf{(2)}The predicted area of each point is given by the filter model, which is compared with the area constraint to remove the architectures that exceed the area constraint. 
% \textbf{(3)}The suggestion model generates the predicted cost of the remaining architectures and sorts the architectures so that the architecture with the best predicted quality can be selected. 
% \textbf{(4)}The sorted architectures are sent to the next steps of the design space exploration flow. 

\section{PIM-Mapper}
\label{sec:mapper}
For a DRAM-PIM accelerator with certain hardware parameters, the performance of DNNs on it depends on the DNN mapping scheme. As illustrated in \Cref{sec:design space}, different DNN layer partition schemes($LM$) result in different communication overhead and computing efficiency in each PIM-node. 
Proper parallelism between \textit{branches}($SM$) can reduce the overhead of layer partitioning while maintaining a balanced load of PIM-nodes. 
The DRAM capacity allocation($WR$) influences the inter-PIM-node communication brought by weight-sharing among PIM-nodes. 
The data layout pattern in DRAM($DL$) affects the cost of DRAM access. 
The effect of each dimension in the mapping space is affected by the other dimensions, so only by considering all the dimensions in the design space of mapping can we get the optimal mapping scheme. 

The PIM-Mapper considers all the aforementioned dimensions of mapping spaces and generates high-hardware-utilization DNN mappings on DRAM-PIM accelerators with given hardware parameters, the flow of which is shown in \Cref{algo:Flow of PIM-Mapper}. 
At the beginning of the optimization process, the PIM-Mapper firstly partitions the input DNN into \textit{segments} as illustrated in \Cref{sec:network level mapping}. 
Considering the constraint on $DL$ between adjacent layers illustrated in \Cref{sec:data layout pattern}, we cannot only optimize the $DL$ for one single layer but have to consider the detailed data dependency between layers of the DNN and choose $DL$ for all layers. 
Thus, we employ an iterative alternated optimization process to optimize the mapping schemes, in which PIM-Mapper firstly optimizes all the $SM$, $LM$ and $WR$ with $DL$ of all layers obtained in the previous iteration and then PIM-Mapper searches for the $DL$ of all layers with the newly solved $SM$, $LM$ and $WR$. 

A dynamic programming algorithm is leveraged for optimizing $SM$, $LM$ and $WR$. As illustrated in \Cref{sec:DRAM capacity allocation}, the value of $WR$ of layers in the DNN is constrained by the DRAM capacity of the PIM-node. 
$WR$ and $LM$ together affect the latency of a layer and the effect of $SM$ is influenced by the $WR$ and $LM$ of the corresponding layers.
So we need to solve $SM$, $LM$ and $WR$ for the whole network simultaneously.
To make full use of the DRAM capacity and explore a variety of $SM$ with different parallelism between \textit{branches}, we firstly generate several candidate $SM$, $LM$ and $WR$ for all \textit{segments} and layers(line\ref{alg_line:candidate generation start}\textasciitilde\ref{alg_line:candidate generation end} in \Cref{algo:Flow of PIM-Mapper}, illustrated in \Cref{sec:mapping candidate generation}). 
Then we select the best combination of the candidates using a dynamic programming algorithm(line\ref{alg_line:candidate selection start}\textasciitilde\ref{alg_line:candidate selection end} in \Cref{algo:Flow of PIM-Mapper}, illustrated in \Cref{sec:mapping candidate selection}). 
\Cref{sec:DL optimization} explains the optimization process of $DL$ of all the layers. 

\begin{algorithm}[htb]
\caption{Flow of the PIM-Mapper}
\label{algo:Flow of PIM-Mapper}
\begin{algorithmic}[1]
\small{
\REQUIRE Configuration of the DRAM-PIM system, a DNN; 
\ENSURE DNN mapping configuration 
\STATE Partition the DNN into $N_{seg}$ \textit{segments}. 
\STATE \textcolor{CommentGreen}{// Initialize mapping schemes}
\FORALL{$n_{seg}\in [1,N_{seg}]$, $l\in[1, L^{n_{seg}}]$} 
\STATE $INIT:\ SM_{obj}^{n_{seg}}, LM_{obj}^{n_{seg},l}, WR_{obj}^{n_{seg},l}, DL_{obj}^{n_{seg},l}$ 
\ENDFOR
\FORALL{$optim\_iter\in [1, MAX\_OPTIM\_ITER]$} 
    \STATE \textcolor{CommentGreen}{// Generate candidates of $SM$, $LM$ and $WR$}\COMMENT{\Cref{sec:mapping candidate generation}} \\ \label{alg_line:candidate generation start} 
    \FORALL{$n_{seg}\in [1, N_{seg}]$}
    \STATE $SM_{can}^{n_{seg}}[1,..,N_{SM}^{n_{seg}}] \gets $ Generate $SM^{n_{seg}}$ with different parallelism of \textit{branches}
    \FORALL{$n_{SM} \in [1, N_{SM}^{seg}]$, $n_{reg} \in [1,N_{seg}^{n_{SM}, n_{reg}}]$, $l \in [1, L^{n_{seg},n_{SM},n_{reg}}]$} 
    \STATE Select the layer with index $(n_{seg}, n_{SM}, n_{reg}, l)$
    \STATE $WR_{can}[1,..,N_{can}]\gets$ Generate different $WR$ values 
    \STATE $LM_{can}[1,..,N_{can}]\gets $ Search for $LM$ with $WR_{can}$
    \STATE Calculate $Perf^{n_{can}}$ and $Size^{n_{can}}$ with $LM_{can}[n_{can}]$ and $WR_{can}[n_{can}]$, $n_{can}\in [1,N_{can}]$
    \ENDFOR
    \ENDFOR \\ \label{alg_line:candidate generation end}
    \STATE \textcolor{CommentGreen}{// Select $SM$, $LM$ and $WR$ from candidates}\COMMENT{\Cref{sec:mapping candidate selection}} \\ \label{alg_line:candidate selection start} 
    \STATE $CS, CL \gets$ MappingSelect($Perf, Size, Cap$) \\ \COMMENT{\Cref{algo:DP-based DRAM capacity allocation}}
    \STATE Update $SM_{obj}^{n_{seg}}$ with $CS[n_{seg}]$, $n_{seg}\in [1, N_{seg}]$
    \STATE Update $LM_{obj}^{n_{seg},l}$ and $WR_{obj}^{n_{seg},l}$ with $CL[n_{seg}][l]$, $n_{seg}\in [1,N_{seg}]$, $l\in[1, L_{n_{seg}}]$ \\ \label{alg_line:candidate selection end} 
    \STATE \textcolor{CommentGreen}{// Update $DL$ with new $LM$ and $WR$} \COMMENT{\Cref{sec:DL optimization}}
    \STATE Search for $DL_{obj}^{n_{seg},l}$ with $LM_{obj}^{n_{seg},l}$ and $WR_{obj}^{n_{seg},l}$, $n_{seg}\in [1,N_{seg}], l\in[1, L_{n_{seg}}]$ \\ \label{alg_line:DL updating} 
\ENDFOR

\RETURN $SM_{obj}$ of each \textit{segment}; $LM_{obj}$, $WR_{obj}$, $DL_{obj}$ of each layer
}
\end{algorithmic}
\end{algorithm}

\subsection{Mapping scheme candidate generation}
\label{sec:mapping candidate generation}
In this step, for each \textit{segment}, we generate candidate $SM$s with different parallelism between \textit{branches}, and for each $SM$ candidate, we generate candidate $LM$-$WR$-pairs with different DRAM capacity requirements for making full use of the DRAM capacity of the PIM-node. 

For each \textit{segment}, we set different values of $N_{reg}$ to get candidate $SM$s with different parallelism between its \textit{branches}. 
For each $N_{reg}$ value, the $IR$ of each \textit{branch} is determined by the number of operations with the objective of balancing the workloads of the $Regions$. 
We leverage the slicing tree representation \cite{lai_slicing_2001} to determine the positions and shapes of the $Regions$, which means we iteratively partition the PIM-node array by two until all the $Regions$ are determined. 
The partitioning process follows the principle of maintaining the size of each $Region$ proportional to the amount of operation of the layers to map so that the PIM-nodes in all $Regions$ can get a balanced load. 
The generated different mapping schemes of the $n_{seg}$-th \textit{segment}, denoted as $SM_{candidate}^{n_{seg}}$, are the candidates from which the final $SM$ of that \textit{segment} is chosen. 

For each candidate $SM$ of a \textit{segment}, we set several $WR$ values for each layer, ranging from the maximum value to 1, to set different DRAM capacity requirements for the layer. 
For each $WR$, we get the corresponding layer mapping scheme $LM$ by traversing all possible $LM$ choices and choosing the one with the lowest latency. 
The different $WR$ values and the corresponding $LM$s form $WR$-$LM$-pairs, and they are the candidates with different latency values and DRAM capacity requirements from which the final $WR$ and $LM$ of the layer are chosen. 
In the first iteration of PIM-Mapper when $DF$ of all layers is not selected yet, we use the amount of DRAM access to estimate the latency of DRAM access and in the remainder iterations, both the amount of access and the data layout pattern are considered.

\subsection{Dynamic-programming-based mapping scheme selection}
\label{sec:mapping candidate selection}
The problem of selecting from the candidate $SM$s of all \textit{segments} and $LM$-$WR$-pairs of all layers is similar to the multiple-choice knapsack problem\cite{kellerer_multiple-choice_2004}, which is to choose exactly one item from each class such that the profit sum is maximized without exceeding the capacity of the knapsack. 
We use a dynamic programming algorithm to solve the mapping scheme selection problem and the flow of the algorithm is shown in \Cref{algo:DP-based DRAM capacity allocation}. 
The inputs of the algorithm are the capacity($CAP$) of DRAM of one PIM-node, different candidate $SM$s denoted as $SM_{can}$, different candidate $LM$-$WR$-pairs of each layer under each $SM_{can}$ and the corresponding latency($Perf$) and required DRAM capacity($Size$) of each layer of each candidate. 
The output of the algorithm is the choice indexes of the candidates. 

We use two tables, $PerfTab$ and $PerfTab_{seg}$, to store the latency with all DRAM capacity values. 
$PerfTab[cap][n_{seg}]$ stores the latency of the first $n_{seg}$ \textit{segments} under the DRAM capacity requirement $cap$. Updated together with $PerfTab$, two tables, $CSTab$ and $CLTab$, are used to store the index of chosen $SM_{can}$ for each \textit{segment} and the index of chosen $LM$-$WR$-pair for each layer, respectively. 
Another table, $PerfTab_{seg}[cap][n_{reg}][l]$ stores the latency result of the first $l$ layers in the $n_{reg}$-th \textit{Region} of a \textit{segment} under the DRAM capacity requirement $cap$. The choice index of $LM$-$WR$-pair of each layer in that \textit{segment} is stored in $CLTab_{seg}$ correspondingly.

Table $PerfTab_{seg}$ is used to collect the latency of layers in one \textit{segment}. For a \textit{segment} with a certain $SM$, we add the latency results of all candidate $LM$-$WR$-pairs of all layers in the \textit{segment} into the $PerfTab_{seg}$ and record the choice index in the $CLTab_{seg}$, which is illustrated in line\ref{alg_line:segment table update start}\textasciitilde\ref{alg_line:segment table update end} in \Cref{algo:DP-based DRAM capacity allocation}. 
During the process, each $LM$-$WR$-pair candidate of each layer in the region is selected to calculate the new latency. 
If the new latency is better than the existing value in the table, that candidate is chosen and the $PerfTab_{seg}$ and $CLTab_{seg}$ are updated. 

Table $PerfTab$ is used to collect the latency of all \textit{segments}. After the latency of a \textit{segment} with all DRAM capacity values is obtained in the $PerfTab_{seg}$, we add the $PerfTab_{seg}$ into $PerfTab$, which is illustrated in line\ref{alg_line:network table update start}\textasciitilde\ref{alg_line:network table update end} in \Cref{algo:DP-based DRAM capacity allocation}. 
The latency result that best improves the total latency under each DRAM capacity value is chosen to update the $PerfTab$, and the $CSTab$ and $CLTab$ are updated correspondingly. 
% The corresponding choice index of the results in the $PerfTab_{seg}$ is stored in the $CLTab_{seg}$. 
% For the $n_{sm}$-th $SM$ of the $n_{seg}$-th \textit{segment}, there are $N_{reg}^{n_{seg}, n_{SM}}$ regions to map the layers and each region has $L^{n_{seg},n_{SM},n_{reg}}$ layers. 
% During the iteration, each $LM$-$WR$-pair candidate of layers in a region is selected to calculate the new latency. 
% If the new latency is better than the existing value in the table, that candidate is chosen and the $PerfTab_{seg}$ and $CLTab_{seg}$ are updated. 
% After getting the $PerfTab_{seg}$ of the $n_{seg}$-th \textit{segment} with the $n_{SM}$-th $SM$, the total latency of that \textit{segment} under each DRAM capacity value can be calculated to be compared with the existing values in the $PerfTab$. 
% The ones that best improves the total latency under all DRAM capacity values are chosen to update the $PerfTab$, and the $CSTab$ and $CLTab$ are updated correspondingly. 

After all $SM$ candidates of all \textit{segments} are visited, the best choices can be acquired in $CSTab[CAP]$ and $CLTab[CAP]$. 

\begin{algorithm}[htb]
\caption{Dynamic Programming for Mapping Selection}
\label{algo:DP-based DRAM capacity allocation}
\begin{algorithmic}[1]
\small{
\REQUIRE The $Perf$ and $Size$ of $N_{can}$ candidate $LM$-$WR$-pairs of each layer in each segment with each candidate $SM$. \ 
The capacity constraint of one PIM-node $CAP$. 
\ENSURE The choice index of $SM_{can}$ for each segment $CS[n_{seg}]$\ 
the choice index of $LM$ and $WR$ for each layer $CL[l^{n_{seg}}]$. 
\STATE \textcolor{CommentGreen}{// Initialize table of $Perf$ and choices}
\STATE $INIT:PerfTab[1,..,CAP][0] = 0$ 
\STATE $INIT:CSTab[1,..,CAP][1,..,N_{seg}]$ 
\STATE {$INIT:CLTab[1,..,CAP][1,..,N_{seg}][1,..,L_{n_{seg}}]$}
\FORALL{$n_{seg} \in [1,N_{seg}]$, $n_{SM} \in [1,N_{SM}^{n_{seg}}]$} 
    \STATE \textcolor{CommentGreen}{// Initialize $table_{seg}$ of $n_{seg}$-th \textit{segment} and $n_{SM}$-th $SM$}
    \STATE $INIT:$\\
    $PerfTab_{\!seg}[1\!,\!..,\!N_{\!reg}^{\!n_{\!seg}\!,\! n_{\!SM}}][1\!,\!..,\!CAP][0] = 0$
    \STATE $INIT: $ \\
    $CLTab_{\!seg}[1\!,\!..,\!N_{\!reg}^{n_{\!seg}\!, \!n_{\!SM}}][1\!,\!..,\!CAP][1\!,\!..,\!L^{n_{\!seg}\!,\!n_{\!SM}\!,\!n_{\!reg}}]$
    \STATE \textcolor{CommentGreen}{// Build the $table_{seg}$ layer-by-layer} \\ \label{alg_line:segment table update start} 
    \FORALL{$n_{reg} \in [1,N_{seg}^{n_{SM}, n_{reg}}]$, $l \in [1, L^{n_{seg},n_{SM},n_{reg}}]$, $cap \in [1, CAP]$, $n_{can} \in [1, N_{can}]$}
    \STATE $Perf_{cur} \gets Perf^{n_{seg},n_{SM},n_{reg},l,n_{can}} + $ \\
    $PerfTab_{seg}[cap][n_{reg}][l-1]$
    \STATE $Size_{cur} \gets Size^{n_{seg},n_{SM},n_{reg},l,n_{can}} + cap$
    \IF{ $Perf_{cur}\leq PerfTab_{seg}[n_{reg}][cap][l]$}
    \STATE $PerfTab_{\!seg}[n_{\!reg}][cap][l][Size_{\!cur}]\!=\!Perf_{\!cur}$
    \STATE Update $CLTab_{seg}$
    \ENDIF
    \ENDFOR \\ \label{alg_line:segment table update end} 

    \STATE \textcolor{CommentGreen}{// Update the table with the  $table_{seg}$} \\ \label{alg_line:network table update start} 
    \FORALL{$cap \in [1, CAP]$, $cap_{seg} \in [1, CAP-cap]$}
    \STATE $Perf_{cur} \gets PerfTab[cap][n_{seg}-1] + $\\
    $\max_{n_{reg}}(\sum_{l}PerfTab_{seg}[cap_{seg}][n_{reg}][l])$
    \STATE $Size_{cur} \gets cap+cap_{seg}$
    \IF {$Perf_{cur}\leq PerfTab[Size_{cur}][n_{seg}]$}
    \STATE $PerfTab[Size_{cur}][n_{seg}] = Perf_{cur}$
    \STATE Update $CLTab[Size_{cur}]$ and $CSTab[Size_{cur}]$ \\ \label{alg_line:network table update end} 
    \ENDIF
    \ENDFOR
\ENDFOR

\RETURN $CSTab[CAP]$ and $CLTab[CAP]$
}

\end{algorithmic}
\end{algorithm}

\subsection{Optimization for data layout pattern}
\label{sec:DL optimization}
With chosen $SM$ for each \textit{segment} and $LM$-$WR$-pairs for all layers, we then update the data layout pattern, $DL$, for each layer. 
Firstly, for each layer with the updated $LM$ and $WR$, we enumerate all possible choices of $DL$ and select the one with the lowest latency without considering the $DL$ of other layers. 
Then we check the $DL$ of each layer pair with data dependency and make the $DL_{i}$ of the successor layer the same as the $DL_{o}$ of the predecessor layer. If the $DL_{i}$ of a layer is changed, we re-select its $DL_{o}$. 

\section{Data-Scheduler}
\label{sec:datascheduler}
In convolution layers and matrix-multiplication layers, there is a large amount of data reuse, and with layer partition methods illustrated in \Cref{sec:layer level mapping}, the temporal reuse of data is converted to spatial data-sharing between PIM-nodes. 
For example, if a layer is partitioned on $K$, the PIM-nodes need to share inputs; if a layer is partitioned on $B$, the PIM-nodes need to share weights. 
To reduce the latency of data-sharing by balancing the load of NoC links, we use a Hamilton-cycle-based data transfer strategy and build an ILP model to determine the Hamilton cycles. 
The data-sharing problem is defined in \Cref{def:data-sharing}. 
Note that a layer can be partitioned on more than one dimension, so there may be multiple \textit{sharing-sets} during one \textit{data-sharing} process. 

\begin{Def}
    \label{def:data-sharing}
    For a piece of data that is stored distributively in a set of PIM-nodes, each PIM-node gets the remaining data from the other PIM-nodes in the set so that eventually the PIM-node has the whole piece of data. The PIM-node set to share data is denoted as a \textit{sharing-set} and the process for the PIM-nodes to get the data is the \textit{data-sharing} process. 
\end{Def}

To achieve a balanced load of PIM-nodes, we use a Hamilton-cycle-based data transfer strategy to schedule the \textit{data-sharing} process: for the PIM-nodes in a \textit{sharing-set} with a Hamilton cycle connecting them, each PIM-node transfers the newly received data to the next PIM-node in the Hamilton cycle and the process is repeated until all PIM-nodes receive all the data. 
This strategy makes all PIM-nodes in the \textit{sharing-set} have equal-sized data to send and receive from NoC. 

In each step of the Hamilton-cycle-based \textit{data-sharing} process, the load of the NoC links is determined by the specific Hamilton cycle thus different Hamilton cycle leads to different data transfer latency. We build an ILP model to determine the Hamilton cycles simultaneously for all \textit{sharing-sets} in one \textit{data-sharing} process. 
For $N_{ss}$ \textit{sharing-sets} where each set has $N_{ns}$ PIM-nodes, we denote the coordinate of each PIM-node as $Coord^{n_{ns},n_{ss}}$. 
We use a binary decision variable $C^{n_{ss},n_{ns}^a,n_{ns}^b}$ to denote the selected connection from $n_{ns}^a$ to $n_{ns}^b$ in the $n_{ss}$-th \textit{sharing-set}. 
The following constraints ensure the selected connections form Hamilton cycles, where integer auxiliary variables $U$ are introduced for eliminating subtours\cite{miller_integer_1960}. 
\begin{equation}
\label{eq:cycle constraint of data-sharing}
\begin{split}
    \sum_{n_{ns}^a=1}^{n_{ns}^a\leq N_{ns}} {C^{n_{ss},n_{ns}^a,n_{ns}^b}} = 1, n_{ns}^b\in[1,N_{ns}], n_{ss}\in[1, N_{ss}]\\
    \sum_{n_{ns}^b=1}^{n_{ns}^b\leq N_{ns}} {C^{n_{ss},n_{ns}^a,n_{ns}^b}} = 1, n_{ns}^a\in[1,N_{ns}], n_{ss}\in[1, N_{ss}]\\
\end{split}
\end{equation}

\begin{equation}
\label{eq:no sub-circle constraint of data-sharing}
\begin{aligned}
    U^{n_{ss}, n_{ns}^a}-U^{n_{ss}, n_{ns}^b}+(N_{ns}-1)\times C^{n_{ss},n_{ns}^a,n_{ns}^b}\leq N_{ns} - 2, \\
    n_{ns}^a, n_{ns}^b\in[2,N_{ns}], n_{ns}^a\neq n_{ns}^b, n_{ss}\in[1,N_{ss}]\\
\end{aligned}
\end{equation}

The latency of data transfer is determined by the link with the heaviest load, so the objective function is to minimize the maximum load of all links in the NoC, which is as follows:
\begin{equation}
\label{eq:objective function of data-sharing}
\begin{aligned}
\footnotesize{
Obj_{ds} \!=\!\max_{Lnk\in\!{Links}}\!\sum_{n_{ss}=1}^{\!n_{\!ss}\!\leq\!N_{\!ss}}\!\sum_{\!n_{\!ns}^a\!,n_{\!ns}^b\!=\!1}^{n_{\!ns}^a\!,n_{\!ns}^b\!\leq\!N_{\!ns}} \!\!\!\!\!Ps(n_{\!ns}^a,\! n_{\!ns}^b,\! Lnk)\! \times\! C^{n_{\!ss}\!,n_{\!ns}^a\!,n_{\!ns}^b}
}
% \begin{array}{lr} C^{n_{\!ss}\!,n_{\!ns}^a\!,n_{\!ns}^b}, &\text{if $n_{ns}^a$ to $n_{ns}^b$ passes Lnk}\\
%      0, &\text{else}
% \end{array}
\end{aligned}
\end{equation}
$Ps(n_{ns}^a, n_{ns}^b, Lnk)$ is 1 if the routing path from PIM-node with index $n_{ns}^a$ to the PIM-node with index $n_{ns}^b$ passes $Lnk$, and otherwise its value is $0$. 

\section{Experiments}
\label{sec:result}
We implement NicePIM on a Linux server with four 18-core Intel Xeon CPUs and four nVidia Tesla V100 GPUs. We use Pytorch\cite{paszke_pytorch_2019} and Botorch\cite{balandat_botorch_2020} to build and train the models of the PIM-Tuner. The PIM-Mapper is implemented using Python language. We use Gurobi\cite{gurobi} to solve the ILP model in the Data-Scheduler.

\subsection{Evaluation methods}
We leverage the DNN accelerator evaluation tool, Timeloop+Accelergy\cite{parashar_timeloop_2019,wu_accelergy_2019}, to get the area of the NN engine of the DRAM-PIM architecture. The intra- and inter-PIM-node DRAM access is simulated by the Ramulator-PIM\cite{kim_ramulator_2016,singh_napel_2019} integrated with BookSim2.0\cite{jiang_detailed_2013}, which are both cycle-accurate simulation tools.  
The DRAMPower\cite{drampower} integrated into Ramulator-PIM helps to get the energy cost of DRAM. 
The latency and energy cost of PE array and buffers for computation tasks are simulated by Timeloop+Accelergy. 

\subsection{Experiment setup}
\label{sec:experiment setup}
The input hardware constraints of NicePIM are shown in \Cref{tab:hardware constraints}. 
The stacked 3D-DRAM has 256 banks with $25nm$ technology node and each bank has $8MiB$ capacity. The energy cost of DRAM access is $0.88pJ/bit$ according to the test result in \cite{fujun_stacked_2020}. 
The DRAM banks are organized into a $16\times 16$ array so that the PIM-node array has a maximum height and width of both 16. 
% The total available area of the logic die excluding the hybrid bonding interface and DRAM Bank controller is set to $48mm^2$ which is inferred from a fabricated PIM chip\cite{niu_184qpsw_2022}. % 5.9mm2 per 256MB. 
The total available area of the logic die for the NN engines is $48mm^2$, which is inferred from a fabricated PIM chip\cite{niu_184qpsw_2022}. % 5.9mm2 per 256MB. 
PIM-nodes run with a clock frequency of 400 MHz and the technology node of logic die is $28nm$.
Each PIM-node can have an up to $256\times 256$ PE array and up to $2048 KiB$ buffers for inputs, weights and outputs. 
The data width of input data and output data of DNN layers is set to 16-bit and the intermediate partial-sums are 32-bit. 
The width of NoC flits is set to half the total width of DRAM banks of a PIM-node and the energy cost is estimated as $1.1 pJ/bit/hop$\cite{wang_ddam_2023}. 
The routers are organized into mesh topology and the dimension-order routing strategy is leveraged with 8 virtual channels. 

The MLP of the filter model of the PIM-Mapper has four layers with 256, 64, 16 and 1 output neurons and the MLP of the suggestion model has three layers with 256, 64 and 16 output neurons. The activation functions of both models are ReLU. 
Adam optimizer\cite{kingma_adam_2017} is leveraged to train the models. 
In each iteration, PIM-Tuner randomly samples architectures from the design space until gets 16384 legal architectures by the filter model. 
% We set a 10\% relaxation when filtering area with the filter model since it is not perfectly accurate. 
The $MAX\_OPTIM\_ITER$ of the PIM-Mapper is set to 3. 

Several DNNs from different fields are used as workloads for evaluation, including GoogLeNet\cite{szegedy_going_2015}, ResNet\cite{he_deep_2016}, VGG\cite{simonyan_very_2015}, DarkNet53\cite{redmon_yolov3_2018} and BERT\cite{devlin_bert_2019}. 
GoogLeNet, VGG16 and ResNet152 are CNNs for classifying images. VGG16 has a straight-line structure while ResNet152 and GoogLeNet have multi-branch structures with short-cut connections and inception-blocks, respectively. 
DarkNet53 is the backbone of the YOLOv3 network used for object detection which has short-cut structures similar to ResNet152. BERT is a kind of Transformer network for natural language processing and we use the BERT-Base model which has 12 heads in one Transformer block. 

% https://www.tablesgenerator.com/
\begin{table}[hbt]
    \centering
    \caption{The hardware constraints}
    \label{tab:hardware constraints}
    \begin{tabular}{l|l|l}
        \hline
        Type                                           & Hardware Parameter        & Value                 \\ \hline
        \multicolumn{1}{c|}{\multirow{5}{*}{Constant}} & Technology node           & $28nm$                \\ \cline{2-3} 
        \multicolumn{1}{c|}{}                          & $BA_{row}\times BA_{col}$ & $16\times 16$         \\ \cline{2-3} 
        \multicolumn{1}{c|}{}                          & $Width_{Bank}$            & $128bit$              \\ \cline{2-3} 
        \multicolumn{1}{c|}{}                          & $CAP_{Bank}$              & $8MiB$                \\ \cline{2-3} 
        \multicolumn{1}{c|}{}                          & $Cstr_{area}$             & $48mm^2$              \\ \hline
        \multirow{7}{*}{Variable}                      & $NA_{row}$                & $2\ \sim\ 16$         \\ \cline{2-3} 
                                                       & $NA_{col}$                & $2\ \sim\ 16$         \\ \cline{2-3} 
                                                       & $PEA_{row}$               & $1\ \sim\ 256$        \\ \cline{2-3} 
                                                       & $PEA_{col}$               & $1\ \sim\ 256$        \\ \cline{2-3} 
                                                       & $Size_{ibuf}$             & $1KiB\ \sim\ 2048KiB$ \\ \cline{2-3} 
                                                       & $Size_{wbuf}$             & $1KiB\ \sim\ 2048KiB$ \\ \cline{2-3} 
                                                       & $Size_{obuf}$             & $1KiB\ \sim\ 2048KiB$ \\ \hline
    \end{tabular}
\end{table}
\subsection{Results of NicePIM}
\begin{figure}[htb]
    \centering
    \includegraphics[width=0.45\textwidth]{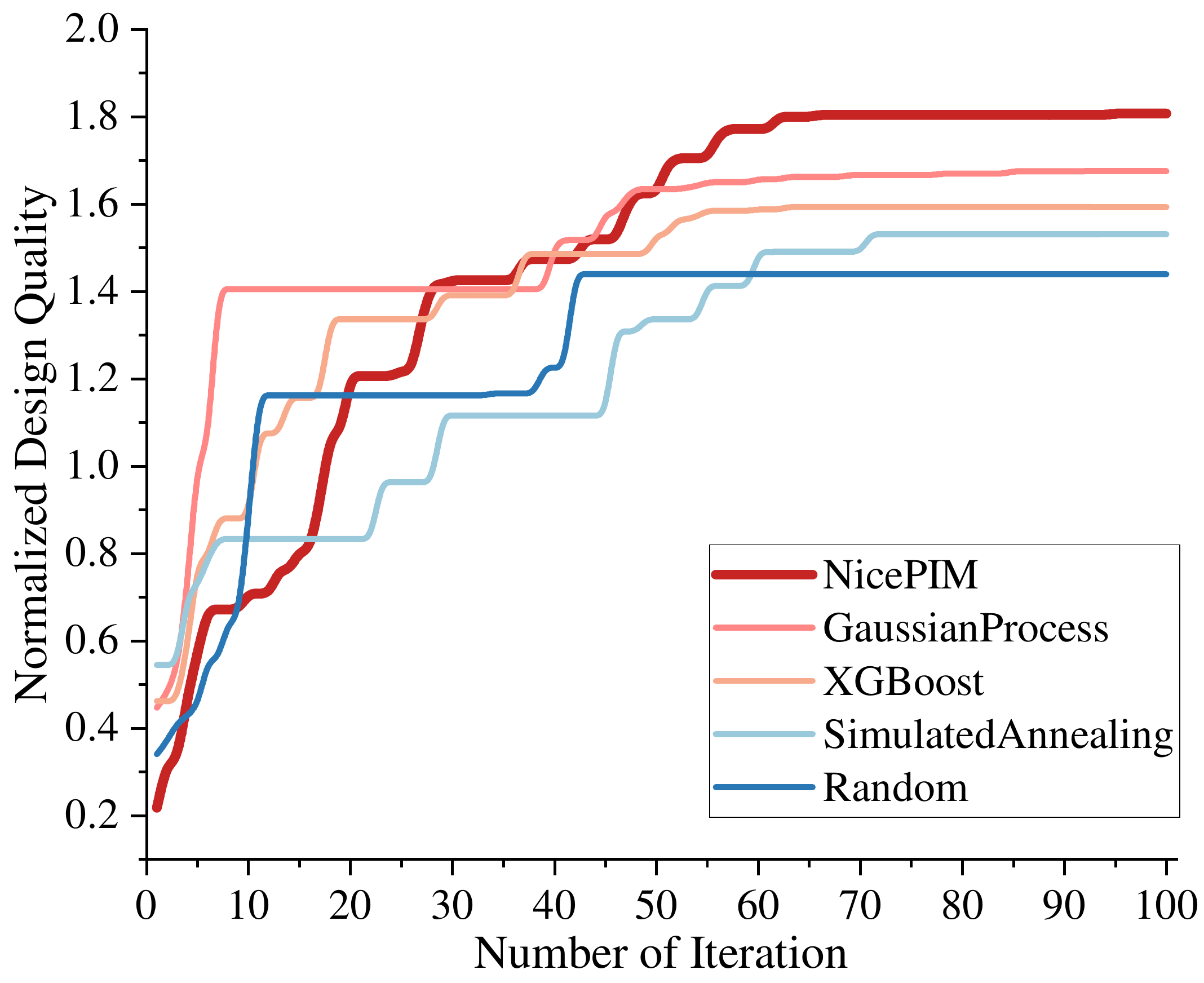}
    \caption{The achieved design quality with NicePIM and other design space exploration methods. The design quality at each iteration is the averaged value of the best three architectures of all evaluated architectures. } 
    \label{fig:dse compare}
\end{figure}

\Cref{fig:dse compare} shows the achieved design quality of NicePIM along with iteration process. 
The optimization goal in \Cref{eq:DesignGoal} is set to $\alpha =1$ and $\beta =1$, which indicates the energy-delay-product(EDP). 
We use the reciprocal of the summed cost of the five DNNs as the metric of the design quality. 

Some other design space exploration methods are evaluated as comparisons, the results of which are shown in \Cref{fig:dse compare}. 
PIM-Mapper and Data-Scheduler are also used with these algorithms for fair comparison. 
In the \textit{Random} method, the architecture to evaluate is randomly chosen in each iteration. 
Another widely used random search algorithm, simulated annealing, is also evaluated. 
Besides, we replace the suggestion model of the PIM-Tuner with other machine learning models. 
In the \textit{GaussianProcess} and \textit{XGBoost} method, the suggestion model is replaced by Gaussian process and XGBoost\cite{chen_xgboost_2016}, respectively. 
The result in \Cref{fig:dse compare} shows that the NicePIM achieves the most significant improvement in design quality. 
The random search algorithms cannot obtain enough information from the already explored architectures while the other two machine learning models are less accurate in characterizing the design space than the suggestion model in the PIM-Mapper. 

Besides, we compare the performance of the nVidia Tesla V100 GPU with the DRAM-PIM architecture given by NicePIM, which has $4\times 8$ PIM-nodes and each PIM-node has a $128\times 8$ PE array with 16KiB, 144KiB and 32KiB buffers for inputs, weights and outputs, respectively. 
% {'array_shape': [4, 8], 'technology_node': 28, 'dram_attr': [128, 64, 8192, 13, 0.88], 'energy_per_flit': 563.2, 'flit_width': 64, 'node_dataflow': 'NVDLA', 'node_arch_config': {'node_shape': (128, 8), 'node_spatial_mapping': [128, 8, 1, 1], 'node_buffer_size': [16384, 147456, 32768], 'access_overlap_factor': 0.0}, 'data_word': [2, 2, 4, 4]}
We try different batch sizes for both systems and choose the best averaged latency per sample as the final performance.  
For DRAM-PIM accelerator, the batch size is changed from 1 to 16, and for GPU, we try batch size from 1 to 1024. 
For fair comparison, we scale the latency results with the area, frequency and technology node. The simulated latency of the DRAM-PIM architecture given by NicePIM is 25x smaller than the tested latency on GPU on average, which means NicePIM makes proper use of the area of the DRAM-PIM system. 

\subsection{Effectiveness of the PIM-Mapper}
\label{sec:experiment mapper}
We evaluate the five DNNs with a batch size of $1$ to illustrate the effectiveness of the PIM-Mapper on two DRAM-PIM systems with $4 \times 4$ and $16 \times 16$ PIM-node arrays. 
In the $4\times 4$ PIM system, we set a $32\times 32$ PE array and $128KiB$ for all SRAM buffers in a PIM-node. As for the $16\times 16$ PIM system, the settings are $8\times 8$ and $8KiB$. 

The results are compared with a baseline method with sequential mapping scheme. 
In baseline method, each layer is mapped onto the whole PIM-node array and we use Timeloop\cite{parashar_timeloop_2019} to solve the $LM$ of the layers with the optimization goal set to ``Delay''. 
The $WR$ of each layer in the baseline method is initially set to the maximum value and if the DRAM capacity is not enough, we iteratively reduce the $WR$ value from the layers with the largest number of weights until the DRAM capacity constraint is met. 
The $DL$ of all layers is set to be the same. We try several $DL$ such as $NCHW$, $NHWC$ and $NCHW[C8]$, and select the one with the best latency result. 
The \textit{data-sharing} process in the baseline method is also scheduled by our proposed Data-Scheduler. 

\begin{figure}[hbt]
    \centering
    \subfloat[\footnotesize{Result on $4 \times 4 $ PIM-node array}]{\label{fig:a}\includegraphics[width=0.48\textwidth]{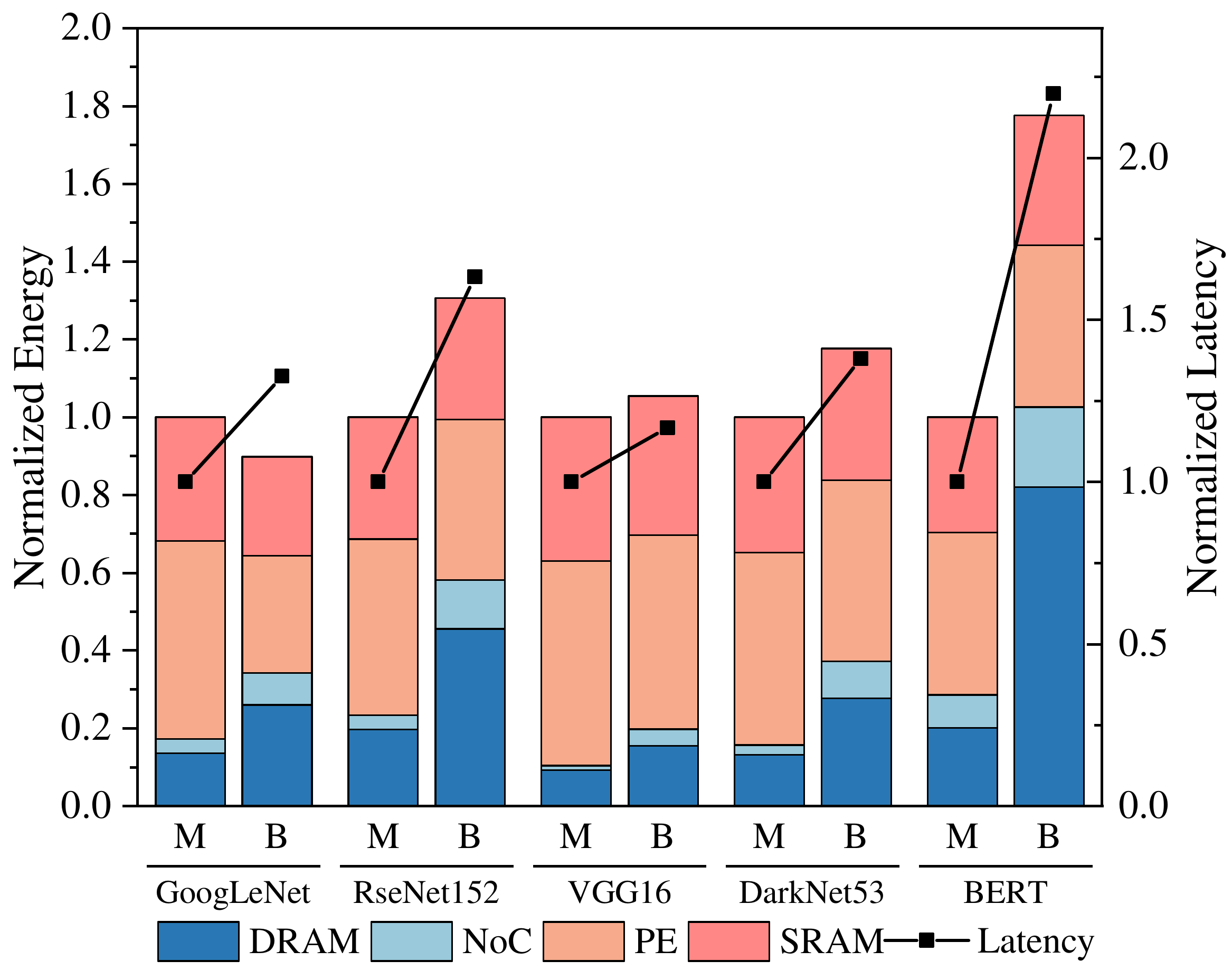}}\\
    \subfloat[\footnotesize{Result on $16\times 16$ PIM-node array}]{\label{fig:b}\includegraphics[width=0.48\textwidth]{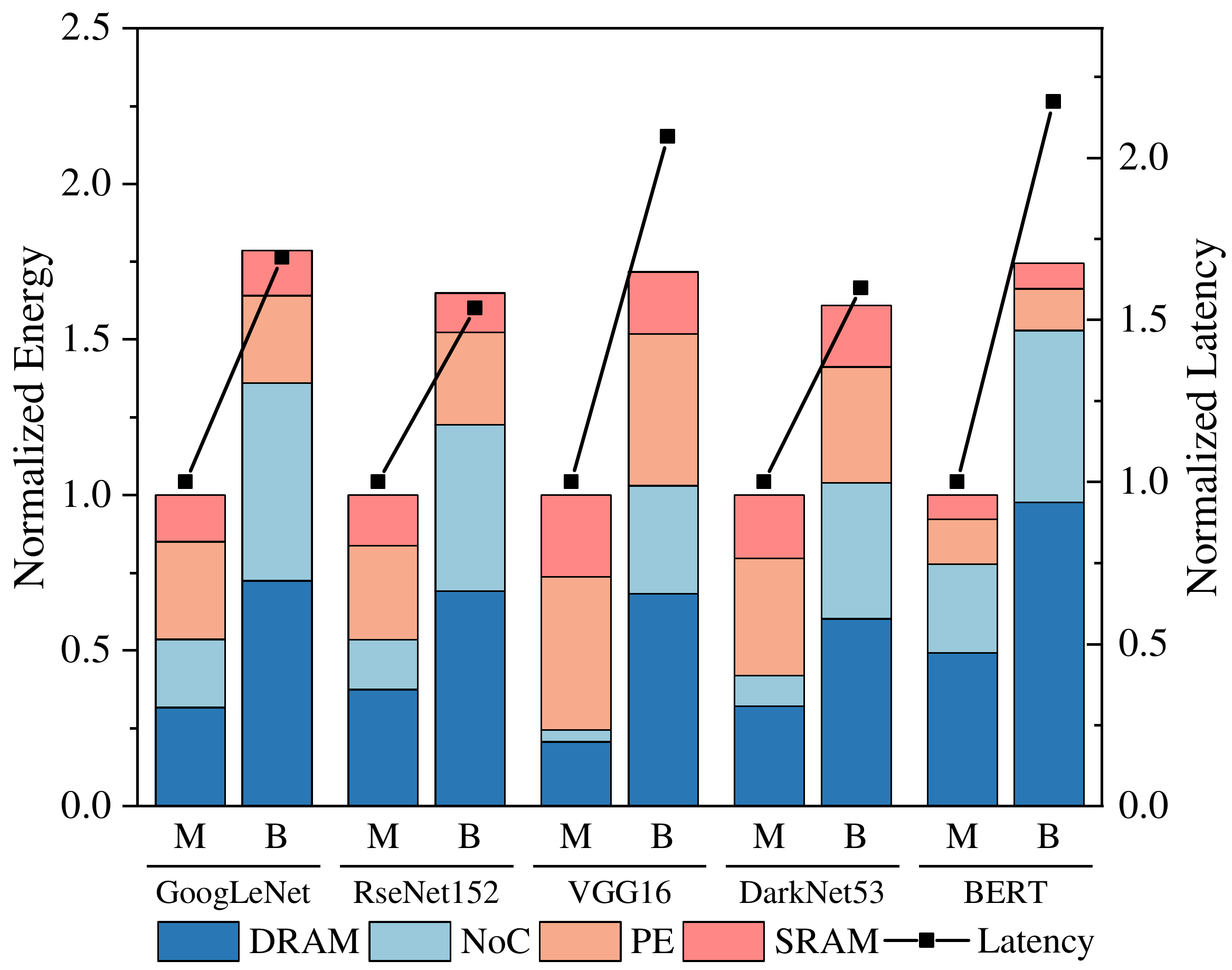}}\\
    \caption{The energy(columns) and latency(lines) of PIM-Mapper(M) and baseline method(B) on DRAM-PIM systems with different-sized PIM-node arrays. The latency and energy results are normalized with that of the PIM-Mapper. }
    \label{fig:result of PIM-Mapper}
\end{figure}

The experimental results in \Cref{fig:result of PIM-Mapper} show that the PIM-Mapper can generate high-utilization and low-energy mappings for PIM systems with different hardware configurations. The latency is reduced by 37\% on average and the energy cost is reduced by 28\% on average. 
On the $4\times 4$ PIM system, where each PIM-node has 16 DRAM banks, the energy cost on DRAM of PIM-Mapper is significantly better than that of the baseline, which means PIM-Mapper better optimizes the $DL$ of layers and makes more sufficient use of the bandwidth of DRAM. 
On the $16\times 16$ PIM system where there are more but smaller PIM-nodes, the energy cost on NoC of PIM-Mapper is much lower than the baseline method, which indicates PIM-Mapper achieves lower inter-PIM-node communication overhead. One reason is that PIM-Mapper better utilizes the parallelism between DNN \textit{branches}. 
Besides, the better DRAM capacity allocation strategy in PIM-Mapper also helps to reduce the overhead for sharing weights. 

\begin{figure}[htb]
    \centering
    \includegraphics[width=3in]{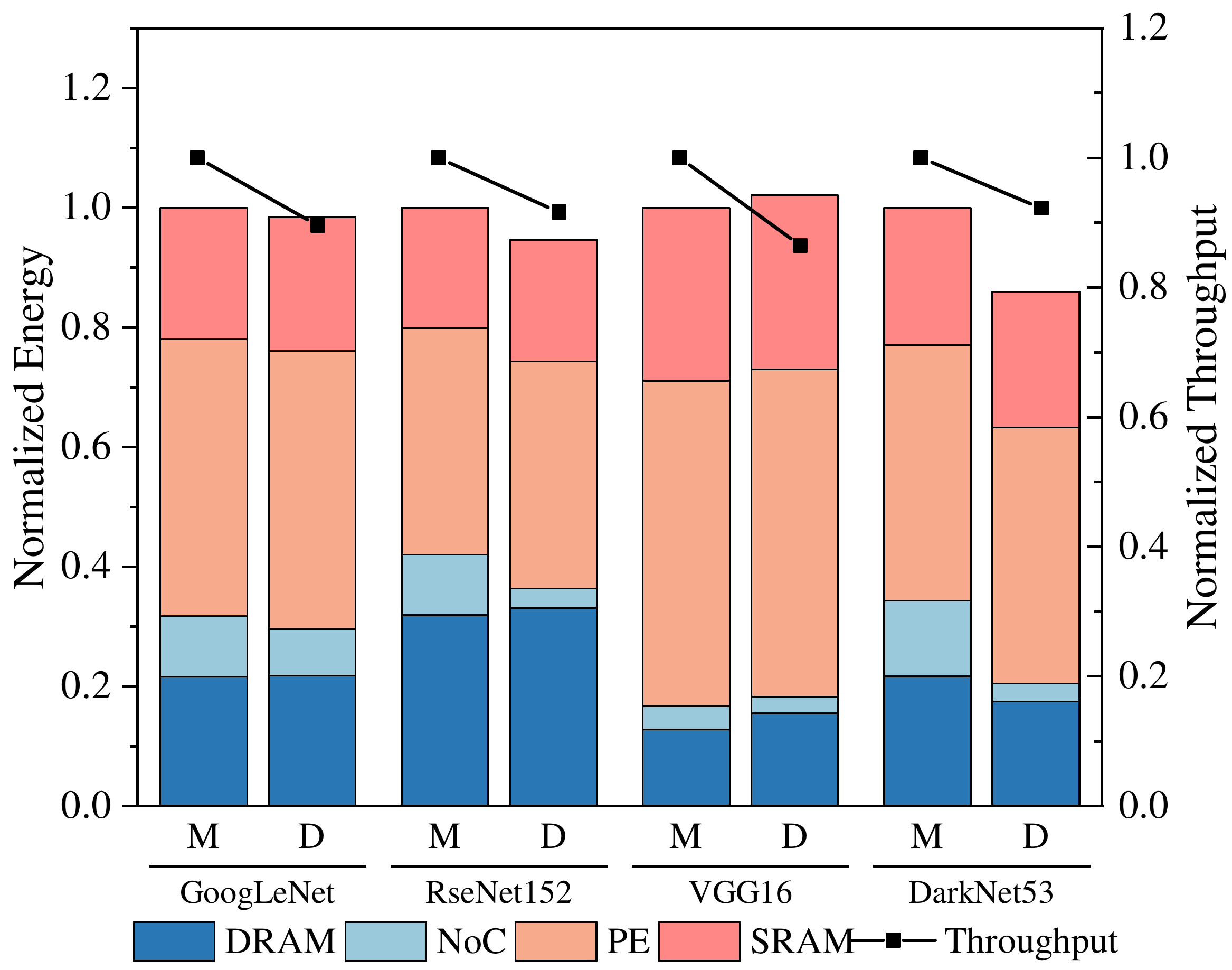}
    \caption{The energy(columns) and throughput(lines) of PIM-Mapper(M) and DDAM(D). The throughput and energy results are normalized to that of the PIM-Mapper.} 
    \label{fig:ddam compare}
\end{figure}

We also compare the PIM-Mapper with DDAM, a CNN mapping framework for DRAM-PIM systems\cite{wang_ddam_2023}. 
DDAM partitions the CNN into several parts and maps them onto different regions of the DRAM-PIM system.
A dynamic programming algorithm is employed to balance the load of the regions of the DRAM-PIM system for high throughput. 
Since DDAM makes CNNs processed in a pipeline manner, we compare the performance on throughput of the two frameworks by changing the batch size from 1 to 16 and choosing the best result. 
The experimental results in \Cref{fig:ddam compare} show that PIM-Mapper achieves better throughput with an average improvement of 11\%. 
Mapping configurations such as data layout pattern and inter-branch parallelism are not taken into account in DDAM which affects its throughput. 
Moreover, DDAM cannot achieve perfect load balance for regions so the utilization of the PIM-node array decreases. 
DDAM and PIM-Mapper have similar energy cost except that of NoC, which is much smaller in the result of DDAM. 
The pipeline-mapping manner leveraged by DDAM can make each layer mapped onto a small region of the DRAM-PIM system and thus the inter-PIM-node communication can be reduced a lot. 
But it is worth noting that the pipeline-mapping scheme employed in DDAM can only be used to optimize the throughput and the latency is 10x worse than PIM-Mapper.  
% while the timestep-by-timestep mapping scheme employed in PIM-Mapper can be used for optimization of both throughput and latency. 
% In some cases, the energy cost of PIM-Mapper is no better than DDAM, this is because the pipeline-mapping scheme employed in DDAM allows each CNN layer to be mapped onto a very small region so that the partitioning of these layers introduces little communication overhead. 
% But it is worth noting that the pipeline-mapping scheme employed in DDAM can only be used to optimize the throughput while the timestep-by-timestep mapping scheme employed in PIM-Mapper can be used for optimization of both throughput and latency. 
% However, the energy of the mapping result of PIM-Mapper is slightly larger than that of DDAM. This is because the pipeline pattern of DDAM allows each CNN layer to be mapped onto a small region so that the communication overhead is small. 

\subsection{Effectiveness of the Data-Scheduler}
\begin{figure}[htb]
    \centering
    \includegraphics[width=3in]{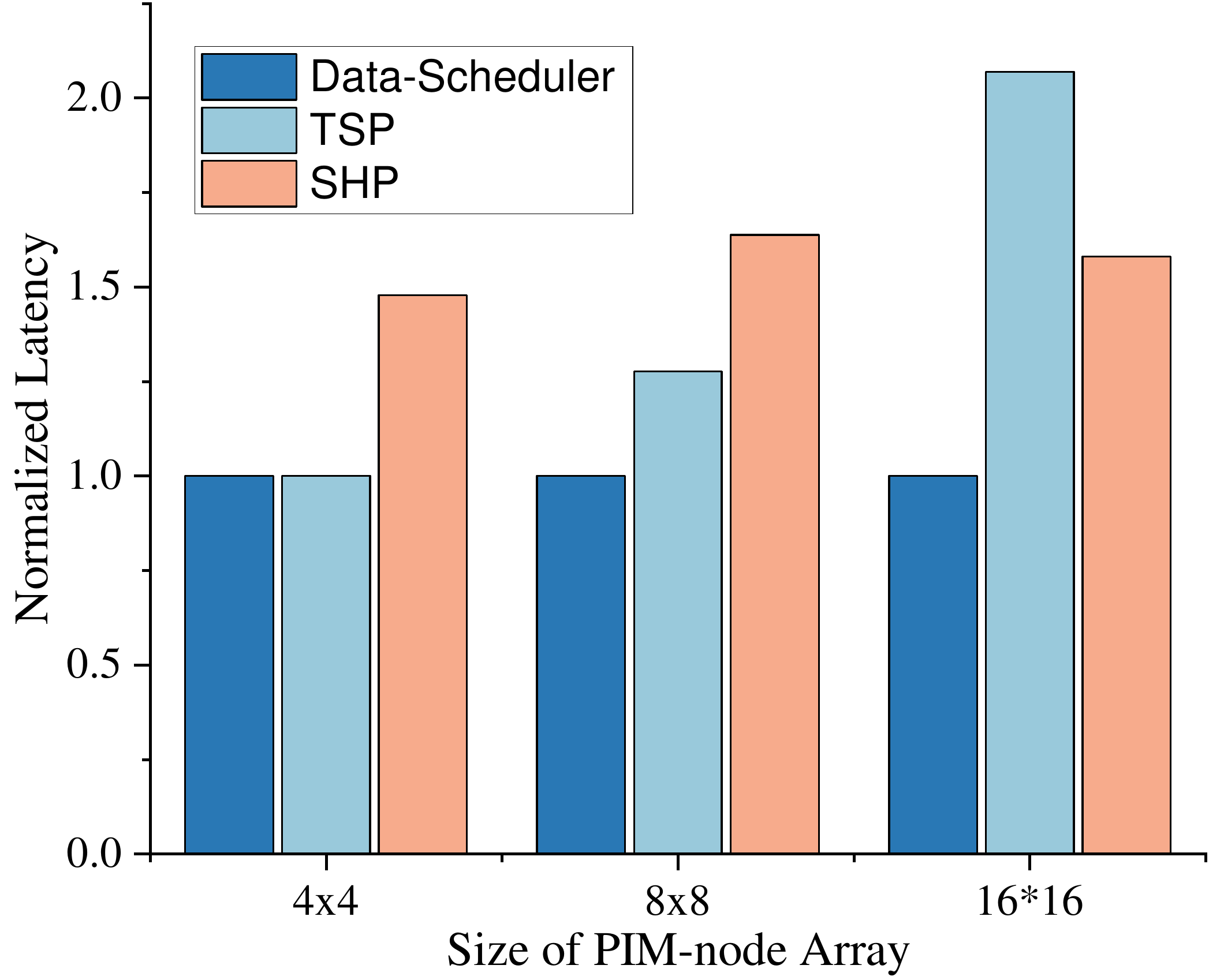}
    \caption{Normalized latency of \textit{data-sharing} on different-sized PIM-node arrays with different scheduling methods. } 
    \label{fig:Data-sharing compare}
\end{figure}

\Cref{fig:Data-sharing compare} shows the comparison of the proposed Data-Scheduler against the other two scheduling methods(\textit{TSP} and \textit{SHP}). 
The method named \textit{TSP} proposed in \cite{wang_ddam_2023} also uses a Hamilton-cycle-based data-transfer pattern while the Hamilton cycle is built by formulating a traveling-sales-man problem. 
The \textit{SHP} method finds the shortest path for each part of the data and then the part-data is transferred along the path, which ensures the smallest hops to transfer all the data. 
We set three sizes for the PIM-node array for evaluation, which are $4\times 4$, $8\times 8$ and $16\times 16$ and the sizes of \textit{sharing set} are all 16. 
On the $8\times 8$ and $16\times 16$ PIM-node array, there are multiple \textit{sharing sets} and they are placed in an interleaving manner: the distances on height and width of adjacent PIM-nodes in the same \textit{sharing set} are all 2 and 4, respectively. For all the PIM-node array sizes, each PIM-node has 8 KiB data to share and the flit width of NoC is 64-bit. 

The results in \Cref{fig:Data-sharing compare} illustrate that the proposed ILP-based scheduling method achieves the smallest latency since the load of links is taken into account. 
The \textit{SHP} method only reduces the hops to transfer data but cannot balance the load of both PIM-nodes and links. 
The \textit{TSP} method also uses the Hamilton path to schedule the data transfer process so that the load of PIM-nodes is balanced. However, the load of links is not taken into account in the \textit{TSP} method, so the latency is still large in some cases. 

\section{Related Work}
\label{sec:related}
\subsection{PIM accelerators with tiled architecture}
Tiled architecture is employed by many 3D-stacking-based PIM accelerators since it has good scalability and matches well with 3D-stacking pattern. 
Kim \textit{et al}. designed a programmable neuromorphic architecture based on Micron's HMC\cite{hmc_2018} named Neurocube\cite{kim_neurocube_2016} as well as a simple mapping strategy that partitions the feature maps of CNNs.  
Gao \textit{et al}. designed TETRIS\cite{gao_tetris_2017}, an HMC-based NN accelerator with data-bypass and in-memory accumulation. TETRIS employs a greedy layer-by-layer partitioning strategy to map CNNs. 
Wang \textit{et al}.\cite{wang_towards_2018} proposed a memory-efficient data allocation strategy for CNNs on 3D-stacked PIM architecture. 
QUEST\cite{ueyoshi_quest_2019} is a 3D-stacked-SRAM-based DNN accelerator and supports log-quantized DNN processing. 
These works mainly focus on the architecture design and scheduling of one PIM-node and use simple DNN mapping strategies. 
DDAM \cite{wang_ddam_2023} is a CNN mapping framework that partitions the CNN into many parts and maps each part onto the different region of the DRAM-PIM system, making the parts processed in a pipeline manner. DDAM can achieve high throughput of CNNs but cannot be used to optimize the latency. 
The hardware design parameters in these aforementioned works and their mapping strategies may not be suitable when the hardware configuration or the target workloads changes. 

\subsection{Design space exploration for DNN accelerators}
The widespread use of DNNs introduces various performance and energy requirements of the accelerators and DNN accelerators have many design parameters to choose. 
Many works are proposed to efficiently explore the design space and find proper design parameters for their target DNN accelerator architectures. 
Timeloop+Accelergy \cite{parashar_timeloop_2019,wu_accelergy_2019} uses a constraint-driven random search method with a fine-grained model for analyzing DNN accelerators to generate valid mappings for DNN layers. 
MAGNet\cite{venkatesan_magnet_2019} has a highly configurable architecture template for DNN accelerators and used Bayesian optimization and random sampling to optimize the hardware configuration, DNN mapping and DNN model. 
ZigZag\cite{mei_zigzag_2021} employs the Memory-Centric Design Space Representation for DNN accelerators and provides heuristic and iterative search strategies to rapidly locate optimal mapping. ZigZag is also able to generate the optimal architecture by exhaustive search. 
FAST\cite{zhang_full-stack_2022} is a framework that jointly explores the hardware datapath configuration, software schedule, and compiler operations for DNN accelerators with detailed DNN performance characterization and a novel op fusion technique. 
To search for effective DNN mapping and efficient hardware configuration, these works have diverse prior definitions on the architecture and DNN mapping, which make them not suitable when facing the design space of DRAM-PIM architectures. 

\section{Conclusion}
\label{sec:conclu}
This paper proposes a framework that optimizes the DNN mapping and hardware parameters for DRAM-PIM-based DNN accelerators. The PIM-Mapper together with the Data-Scheduler can effectively reduce the inference latency and the energy cost of DNNs on DRAM-PIM architectures with various hardware parameters. The PIM-Tuner is effective to extract features from the hardware design space so that the obtained architecture has higher quality compared to other design space exploration methods.

% \section*{Acknowledgments}
% The authors would like to thank Information Science Laboratory Center of USTC for the hardware \& software services.

\bibliographystyle{IEEEtran}
% \bibliography{./ref/bstcontrol,./ref/PIMPaper,./ref/website}

\newpage

\iffalse
\section{Biography Section}
If you have an EPS/PDF photo (graphicx package needed), extra braces are
 needed around the contents of the optional argument to biography to prevent
 the LaTeX parser from getting confused when it sees the complicated
 $\backslash${\tt{includegraphics}} command within an optional argument. (You can create
 your own custom macro containing the $\backslash${\tt{includegraphics}} command to make things
 simpler here.)
 
\vspace{11pt}

% Photos in the template
\bf{If you include a photo:}\vspace{-33pt}
\begin{IEEEbiography}[{\includegraphics[width=1in,height=1.25in,clip,keepaspectratio]{fig1}}]{Michael Shell}
Use $\backslash${\tt{begin\{IEEEbiography\}}} and then for the 1st argument use $\backslash${\tt{includegraphics}} to declare and link the author photo.
Use the author name as the 3rd argument followed by the biography text.
\end{IEEEbiography}

\vspace{11pt}

\bf{If you will not include a photo:}\vspace{-33pt}
\begin{IEEEbiographynophoto}{John Doe}
Use $\backslash${\tt{begin\{IEEEbiographynophoto\}}} and the author name as the argument followed by the biography text.
\end{IEEEbiographynophoto}

\vfill
\fi

\end{document}